\documentclass[letterpaper,12pt]{article}
\usepackage{authblk}
\usepackage[utf8]{inputenc}
\usepackage[english]{babel}
\usepackage{amsmath}
\usepackage{amssymb}
\usepackage{graphicx}
\usepackage{multirow}
\usepackage{epsfig}
\usepackage{rotating}
\usepackage{longtable} 
\usepackage{caption}
\usepackage{subcaption}
\usepackage{url}
\usepackage{pdflscape}
\usepackage{lscape}
\usepackage{cancel}
\usepackage{setspace}
\usepackage{csquotes}
\usepackage{placeins}
\usepackage[left=2cm,top=3cm,right=2cm,bottom=3cm]{geometry}
\usepackage{tabularx}
\usepackage{adjustbox}
\usepackage{color,xcolor}
\usepackage{comment}
\usepackage{tensor}
\usepackage{hyperref}

\usepackage{mathtools}
\usepackage{latexsym}

\title{Higgs-like Resonances and Massive Neutrinos in a 3-3-1 Model}

\author[a]{Richard H. Benavides\thanks{richardbenavides@itm.edu.co}}
\author[b]{D.V. Forero\thanks{dvanegas@udemedellin.edu.co}}
\author[c]{Eduardo Rojas\thanks{eduro4000@gmail.com}}
\author[b]{A. Tapia\thanks{atapia@udemedellin.edu.co}}

\affil[a]{Instituto Tecnol\'{o}gico Metropolitano, Facultad de Ciencias Exactas y Aplicadas,  \textit{\small Calle 73 N° 76-354 vía el volador, Medell\'{i}n, Colombia.}}

\affil[b]{Instituto de Ciencias Básicas, Universidad de Medellín,  \textit{\small Carrera 87 N° 30–65, Medellín, Colombia.}}

\affil[c]{Departamento de Física, Universidad de Nariño, \textit{\small Calle 18 Carrera 50, A.A. 1175, San Juan de Pasto, Colombia}}

\date{\today}

\begin{document}

\maketitle

\begin{abstract}
Recent experimental results have reported mild deviations from Standard Model predictions in processes involving two photons in the final state, suggesting the possible presence of high-mass scalar resonances at the few-hundred-GeV scale. We investigate these anomalies within the framework of 3-3-1 models, a well-motivated class of extensions of the Standard Model. Focusing on the most relevant regions of parameter space, we determine the preferred scalar mass ranges and present the results in terms of probability density functions. We implement the 3-3-1 model with right-handed neutrinos, which can be considered as a benchmark within this class of models, in the \texttt{SARAH} package. The scalar sector is constructed from the most general potential involving three Higgs triplets and one scalar sextet, consistent with the required symmetries for a realistic model. The neutrino sector is also analyzed in some detail, where left- and right-handed neutrinos reside in the same leptonic multiplet. Neutrino masses are generated via the Type-I seesaw mechanism, and the Yukawa couplings were constrained employing the Casas–Ibarra parametrization. Finally, we discuss the constraints arising from flavor-changing neutral currents and electroweak precision data.
\end{abstract}

\section{Introduction}
The Standard Model (SM) of particle physics provides a remarkably successful description of the fundamental interactions among known particles. Despite its achievements, the SM is incomplete: it cannot account for the origin of neutrino masses, the replication of fermion families, or the existence of dark matter, among other open problems~\cite{dms,ParticleDataGroup:2024cfk,Mohapatra:2005wg,Bertone:2004pz}. These shortcomings strongly motivate the study of extensions of the gauge and fermionic sectors capable of addressing such phenomena.

Collider experiments are also actively probing new physics through direct searches. In particular, extended scalar sectors are of special interest, since they often give rise to new resonances that could appear in precision measurements or in high-energy final states. Recent analyses by the ATLAS and CMS Collaborations have reported mild but intriguing deviations from SM predictions in diphoton channels at invariant masses of a few hundred GeV~\cite{ATLAS:2018mgv,CMS:2018dqv}. While not yet statistically conclusive, these anomalies can be interpreted as possible hints of new scalar resonances, motivating their study within theoretically consistent frameworks. In this work, we examine these anomalies in the context of the 3-3-1 model with right-handed neutrinos, scanning the most motivated regions of parameter space and identifying the preferred mass ranges of the additional scalar states. In particular, we considered the case of an excess near 95~GeV (diphoton resonance), that has a combined/global significance of about $3.8\sigma$, a long–standing low-mass diphoton hint, consistently appearing in CMS data and supported by LEP and ATLAS observations~\cite{Ashanujjaman:2023etj}.\\

On the other hand, the observation of the neutrino flavor conversion in solar neutrinos, and the observation of neutrino oscillations from different neutrino sources~\cite{Koshio:2022zip} in general, have provided direct experimental evidence for physics Beyond the SM~\cite{Arguelles:2022tki,deGouvea:2022gut}. Therefore, a realistic model must account for this phenomenology, in particular what concerns with the active neutrino masses and mixings~\cite{deSalas:2020pgw}. Although the mechanism behind the generation of the neutrino masses is unknown, several realizations of the dimension five operator are possible at three level as well as at the loop level, providing the correct mass-scale for the active neutrinos. In the context of the 3-3-1 models, massive neutrinos have been included in the model at tree-level~\cite{Valle:1983dk,Tully:2000kk,Montero:2001ts} as well as at the loop-level\cite{Okada:2015bxa}. In the specific case of models with a scalar sextet see for instance Ref.~\cite{Ky:2005yq} and Refs.~\cite{Montero:2000rh,Dong:2008sw, Pinheiro:2022obu} for the implementation of the Type-I and Type-II seesaw mechanisms, respectively. In this work, we considered a 3-3-1 model with right-handed neutrinos, without exotic charges, where the Type-I seesaw mechanism is implemented by the addition of a scalar sextet.


The 3-3-1 models, based on the gauge group $SU(3)_c \times SU(3)_L \times U(1)_X$, form an attractive class of SM extensions. In particular, the non-universal realizations of the gauge symmetry~\cite{Pisano:1991ee,Frampton:1992wt,Foot:1994vd,Ponce:2001jn,Benavides:2021pqx} offer appealing theoretical features: it explains why the number of fermion families must be three (or a multiple of three under asymptotic freedom), and it naturally accommodates an extended scalar sector typically built from three Higgs triplets~\cite{Long:1997vbr}. In our study, we consider an enlarged scalar sector, where the addition of a scalar sextet provides the necessary structure for a realistic mass spectrum in the scalar, lepton, and quark sectors. \\

The structure of this paper is as follows. In Sec.~\ref{sec:general331} we review the main features of 3-3-1 models without exotic electric charges. Sec.~\ref{sec:331RHN} provides a brief overview of the specific realization with right-handed neutrinos, which serves as our benchmark scenario. In Sec.~\ref{subsec:escalar} we construct the most general scalar potential consistent with the required symmetries. This potential yields a realistic spectrum with the correct number of Goldstone bosons and a phenomenologically viable set of physical scalars that can be confronted with collider anomalies, thereby establishing a direct link between theory and experiment.  
In Sec.~\ref{sec:analisis-analitico} we describe the analytical implementation. The model is first implemented in \verb|SARAH| package within \verb|Mathematica|, which generates the full set of model files and allows us to validate the analytical results.  
The results of the scalar spectrum analysis are presented in Sec.~\ref{sec:pdf}.
In Sec.~\ref{sec:seesaw} we turn to the neutrino sector. Since left- and right-handed neutrinos belong to the same leptonic multiplet in this model, the inclusion of a scalar sextet allows the simultaneous generation of Dirac and Majorana mass terms~\cite{Ky:2005hep,DeConto:2015}. Light neutrino masses are then obtained via a Type-I seesaw mechanism, and we employ the Casas–Ibarra parametrization~\cite{Casas:2001sr} to extract the corresponding Yukawa couplings numerically.  
Finally, in Sec.~\ref{sec:ew-fcnc} we analyze the phenomenological constraints arising from electroweak precision data and from flavor-changing neutral currents, which further restrict the viable parameter space of the model.  

\section{General Aspects of \texorpdfstring{$SU(3)_c\otimes SU(3)_L\otimes U(1)_X$}{} Models\label{sec:general331}}

Since the Standard Model (SM) cannot be considered a complete theory, various extensions have been proposed to address phenomena it fails to explain, while still reproducing its predictions at low energies. The most straightforward approach is to extend the SM itself. This can be achieved by enlarging the fermionic sector~\footnote{For instance, by simply introducing right-handed neutrinos into the SM, it is possible to generate either Dirac or Majorana masses for neutrinos, thus providing a framework to explain the experimental observations related to neutrino oscillations.}, increasing the scalar sector with more than one Higgs representation, or extending the local gauge group.

In this direction, one theoretical framework is based on the local gauge symmetry $SU(3)_c\otimes SU(3)_L\otimes U(1)_X$ commonly referred to as the 3-3-1 model for short~\cite{maha,Martinez,Sanchez,Fanchiotti,Pleitez,florez}. A key motivation for these models is their ability to explain the number of fermion families observed in nature an open question that the SM cannot address, as it imposes no constraint on the number of generations, aside from the upper bound $N\leq 8$ arising from the requirement of asymptotic freedom in Quantum Chromodynamics (QCD)~\cite{dms}. Experimental results from CERN–LEP in the 90's indicate the existence of at least three families, each containing a neutral lepton with a mass less than half the mass of the neutral gauge boson $Z$. The 3-3-1 models naturally explain why the number of fermion families must be three by requiring anomaly cancellation in the gauge current algebra. This cancellation is possible because, in these models, the first two families have different quantum numbers from the third. Nevertheless, LEP data do not exclude the existence of additional families, as long as their associated neutral leptons (neutrinos) are heavy~\footnote{That is, the mass of the new neutral leptonic fields must be greater than half the mass of the neutral gauge boson 
$Z$.}. The 3-3-1 framework can also accommodate the existence of more fermion families under certain theoretical conditions.\\

\noindent The main features of the 3-3-1 models are:

\begin{itemize}
 \item In some of these models, the gauge anomalies are canceled only if the number of families is a multiple of three (to ensure the asymptotic freedom of $SU(3)_c$)  \cite{Pleitez},
\cite{Pisano}, \cite{Frampton}, \cite{ozer}.
\item The Peccei–Quinn symmetry can be easily implemented \cite{Peccei}, \cite{Dias}.
\item One quark family has different quantum numbers, a feature that can be used to explain the large mass of the top quark \cite{Fanchiotti}, \cite{Frampton:1995wf}.
\item The scalar sector contains good dark matter candidates
\cite{Fregolente:2002nx}, \cite{Hoang:2003vj}, \cite{Filippi:2005mt}.
\item The leptonic sector is capable of describing certain properties of neutrinos \cite{Okamoto:1999cf}, \cite{Kitabayashi:2000nq}, \cite{Chang:2006aa}.
\item The hierarchy in the Yukawa coupling constants can be avoided by implementing the universal see-saw mechanism. \cite{Fanchiotti}, \cite{Davidson:1987mh}, \cite{Rajpoot:1987ji}, \cite{Chang:1986bp}, \cite{Gutierrez:2005rq}.
\end{itemize}

\noindent The above features make these models highly attractive for theoretical and phenomenological studies. It is assumed that the electroweak gauge group $SU(3)_L\otimes U(1)_X \supset SU(2)_L \otimes U(1)_Y$,
is a symmetry extension of the SM, where the left-handed quarks are color triplets and the right-handed leptons are color singlets. These fermions transform under the two fundamental representations, $(3  $ y $ 3^\ast)$ of $SU(3)_L$. \\

In 3-3-1 models the charge operator is given by,

\begin{equation}\label{eq21}
Q=aT_{3L}+\frac{2}{\sqrt{3}}bT_{8L}+XI_3,
\end{equation}

\noindent where $T_{iL}=\lambda_{iL}/2$; $\lambda_{iL}$ are 
Gell-Mann matrices for $SU(3)_L$, normalized as $Tr(\lambda_i
\lambda_j=2\delta_{ij})$; $I_3=\text{diag}(1,1,1)$ is the unit matrix
$3 \times 3$; $a$ and $b$ are arbitrary parameters. By setting $a=1$ one recovers the usual isospin of the electroweak interaction in the SM, thereby fixing the value of $a$.\\

\noindent In principle, there are infinitely many possible values for the parameter $b$, which leads to an infinite number of 3-3-1 models. Taking into account this, the eight gauge fields can be written in the form

\begin{equation}\label{gfi}
\sum_\alpha\lambda_\alpha A^\alpha_\mu=\sqrt{2}\left(
\begin{array}{ccc}
D^0_{1\mu} & W^+_\mu & K_\mu^{(b+1/2)} \\
W^-_\mu & D^0_{2\mu} & K_\mu^{(b-1/2)} \\
K_\mu^{-(b+1/2)} & K_\mu^{-(b-1/2)} & D^0_{3\mu} \\
\end{array}\right),
\end{equation}

\noindent where $D^0_{1\mu}=A_\mu^3/\sqrt{2}+A_\mu^8/\sqrt{6},\; 
D^0_{2\mu}=-A_\mu^3/\sqrt{2}+A_\mu^8/\sqrt{6},$ y 
$D^0_{3\mu}= -2A_\mu^8/\sqrt{6}.$  The superscripts on the gauge bosons determine the electric charge of the particles, which depends solely on the value of the parameter $b$. In order to obtain integer electric charges, $b$ must take the values
 $b=\pm 1/2$, $\pm 3/2$, $\pm 5/2$,..., $\pm(2n+1)/2$,
$n=1,2,3...$. However, the negative values can be disregarded since they are related to the positive ones by taking the complex conjugate of the covariant derivative, which is equivalent to replacing $(3\rightarrow 3^*) $ in the fermionic content. The electric charge operator defined in Eq.~(\ref{eq21}), under the $(3$ y $3^*)$ $SU(3)_L$ representations can be written as~\cite{Ponce:2002sg}

\begin{eqnarray}\label{carga}
Q[3]=\text{diag}\left(\frac{1}{2}+\frac{b}{3}+X,\; -\frac{1}{2}+\frac{b}{3}+X,\; -\frac{2b}{3}+X\right),\\ \nonumber
Q[3^*]=\text{diag}\left(-\frac{1}{2}-\frac{b}{3}+X,\; \frac{1}{2}-\frac{b}{3}+X,\; \frac{2b}{3}+X\right). \nonumber
\end{eqnarray}

\noindent The values of $X$ are determined by the cancellation of triangular anomalies, which for these models are given by $[SU(3)_L]^3$, $[SU(3)_c]^2U(1)_X$,  $[SU(3)_L]^2U(1)_X$,  $[\text{grav}]^2U(1)_X$ y  $[U(1)_X]^3$. From these relations, the values of  $X$ can be determined once the value of $b$ is fixed.

\section{The 3-3-1 Model With Right-Handed Neutrinos\label{sec:331RHN}}
The requirement of avoiding fields with exotic electric charges is satisfied by taking 
\(b=\tfrac{1}{2}\) in Eq.~(\ref{eq21}), as shown in Ref.~\cite{Ponce:2001jn}. 
An important realization of this choice is the so-called 3-3-1 model with right-handed 
neutrinos, which constitutes the focus of the present work. 
Originally proposed by Foot, Long and Tran~\cite{Foot:1994ym}, this version of the model 
arranges each lepton family into a triplet containing the left-handed neutrino, the charged 
lepton, and the charge-conjugated right-handed neutrino, thereby incorporating \(\nu_R\) 
naturally into the gauge structure. 
In contrast to the minimal 3-3-1 model~\cite{Pleitez}, this formulation requires only three scalar triplets to implement the spontaneous symmetry breaking, while still providing a simple framework for neutrino mass generation~\cite{Long:1995ctv}. 
In the present study we also extend the scalar sector by introducing a sextet, which enables the implementation of a seesaw mechanism for neutrino masses.

\subsection{ The Fermionic Sector}\label{subsec:fermionico}

Quarks can be represented in the model in the form $Q_L^i=(u^i,d^i,D^i)_L\sim (3,3,0)$, for the first two families $i=1,2$, where $D^i_L$ refers to two additional quarks with electric charge $-1/3$. The quantum numbers in parentheses correspond to the symmetries $[SU(3)_c, SU(3)_L,U(1)_X]$, $Q_L^3=(d^3,u^3,U^3)_L\sim (3,3^*,1/3)$, where $U_L$ is an extra up-type quark. Right-handed quarks are represented in the form $u_L^{ac}\sim (3^*,1,-2/3)$, $d_L^{ac}\sim (3^*,1,1/3)$ with $a=1,2,3$ family index, $D_L^{ic}\sim (3^*,1,1/3)$, $i=1,2$ y $U_L^c \sim (3^*,1,-2/3)$.\\

Leptons can be represented in the model in the form $L_L^l=(l^-,\nu_l^0,\nu_l^{0c})_L\sim (1,3^*,-1/3)$, for  $l=e,\mu,\tau$, leptonic family index, and the three singlets  $l_L^+\sim (1,1,1)$ where $\nu_l^0$ is the field associated with the neutrino and $\nu_l^{0c}$ it plays the role of a right-handed neutrino. This is a gauge anomaly free model, in which all three right handed neutrinos are embedded within the same leptonic multiplet, a feature that gives the model a name for reference.

\subsection{The Scalar Sector \label{subsec:escalar}}

The scalar sector within the framework of the 3-3-1
 models without exotic electric charges that we will consider consist of three scalar triplets and one scalar sextet, as presented in Refs.~\cite{Ky:2005yq,Roberto,Pinheiro:2022obu,Tully:1998wa}:

\begin{equation}
  \Phi_1(1,3,1/3)=\begin{pmatrix} \phi_1^0\\\phi_1^- \\\phi_1^{\prime 0}\end{pmatrix}, \,\,\,\text{with VEV:}\,\,\langle \Phi_1 \rangle=\frac{1}{\sqrt{2}}\begin{pmatrix} V_1\\ 0 \\ v_1\end{pmatrix},
\end{equation}

\begin{equation}
  \Phi_2(1,3,1/3)=\begin{pmatrix} \phi_2^0\\\phi_2^- \\\phi_2^{\prime 0}\end{pmatrix}, \,\,\,\text{with VEV:}\,\,\langle \Phi_2 \rangle=\frac{1}{\sqrt{2}}\begin{pmatrix} v_2\\ 0 \\ V_2\end{pmatrix},
\end{equation}

\begin{equation}
  \Phi_3(1,3,-2/3)=\begin{pmatrix} \phi_3^+\\\phi_3^0 \\\phi_3^{\prime +}\end{pmatrix}, \,\,\,\text{with  VEV:}\,\,\langle \Phi_3 \rangle=\frac{1}{\sqrt{2}}\begin{pmatrix} 0 \\ v_3 \\ 0\end{pmatrix},
\end{equation}

\begin{equation}
  \Phi_4(1,6,2/3)=\frac{1}{\sqrt{2}}\begin{pmatrix}\sqrt{2} \phi_4^0  & \phi_4^+ & \phi_4^{\prime 0} \\ \phi_4^+ &  \sqrt{2} \phi_4^{++} &  \phi_4^{\prime +} \\ 
 \phi_4^{\prime 0}   & \phi_4^{\prime+} & \sqrt{2}\phi_4^{\prime\prime 0}\end{pmatrix}, \,\,\,\text{with VEVs:}\,\,\langle \Phi_4 \rangle=\begin{pmatrix} W_1& 0 & W_2/\sqrt{2} \\ 0 & 0 & 0  \\ W_2 /\sqrt{2}  &  0 & W_3 \end{pmatrix}.
\end{equation}

With the hierarchy  $ V_1  \sim v_3 \sim 10^2\,\, \text{GeV}$, such that $v_{SM} =\sqrt{V_1^2+v_3^2} = 246$~GeV  and $V_2 \approx 6\,\text{TeV} \gg V_1, v_3$. In addition, a vacuum alignment has been performed in such a way that $v_1=0$ and $v_2=0$. We also assume the hierarchy $W_1 \ll W_2 \ll W_3$ in order the be compatible with the Type-I seesaw~\cite{Ky:2005yq}, and for simplicity we will set $W_1 = 0$.\\

From the above scalar content, the most general scalar potential allowed by the symmetries of the model, contains the following terms:

\begin{align}\label{eq:Potencial}
    V(&\Phi_1,\Phi_2,\Phi_3,\Phi_4)\notag\\ 
    & = \mu^2_{\Phi_1} \Phi_1^\dagger\cdot \Phi_1 +\mu^2_{\Phi_2} \Phi_2^\dagger\cdot\Phi_2 + \mu^2_{\Phi_3} \Phi_3^\dagger\cdot \Phi_3+\mu^2_{\Phi12}\left(\Phi_2^{\dagger}\cdot\Phi_1+ h.c.\right) + \mu^2_{\Phi_4} \text{Tr}[\Phi_4^\dagger \cdot \Phi_4]\nonumber \\
    & + \lambda_1 (\Phi_1^\dagger\cdot\Phi_1 )^2+ \lambda_2 (\Phi_2^\dagger\cdot\Phi_2)^2 +\lambda_3 (\Phi_3^\dagger\cdot\Phi_3)^ 2\nonumber \\
    & +\left(\lambda_4 (\Phi_1^\dagger \cdot \Phi_1)(\Phi_2^\dagger \cdot \Phi_2)+ h.c.\right) + \lambda_5 (\Phi_1^\dagger \cdot \Phi_1)(\Phi_3^\dagger \cdot \Phi_3) \nonumber
    \\ & 
     +\left( \lambda_6 (\Phi_2^\dagger \cdot \Phi_1)(\Phi_3^\dagger \cdot\Phi_3) + h.c.\right)+\lambda_7 (\Phi_2^\dagger \cdot \Phi_2)(\Phi_3^\dagger \cdot \Phi_3) 
    + \lambda_8 (\Phi_1^\dagger \cdot \Phi_2)(\Phi_2^\dagger \cdot \Phi_1)
    \nonumber
    \\& 
     + \lambda_9 (\Phi_1^\dagger \cdot \Phi_3)(\Phi_3^\dagger \cdot \Phi_1) + \lambda_{10} (\Phi_2^\dagger \cdot \Phi_3)(\Phi_3^\dagger \cdot \Phi_2) + \left(\lambda_{11} (\Phi_2^\dagger \cdot \Phi_3)(\Phi_3^\dagger \cdot \Phi_1)+ h.c.\right)
     \nonumber 
     \\ & 
    +\lambda_{12} (\Phi_3^\dagger \cdot \Phi_4\cdot \Phi_4^\dagger \cdot \Phi_3) +  \lambda_{13} (\Phi_1^\dagger \cdot \Phi_4\cdot \Phi_4^\dagger \cdot \Phi_1) + 
    \left(\lambda_{14} (\Phi_1^\dagger \cdot \Phi_4\cdot \Phi_4^\dagger\cdot \Phi_2) + h.c.\right)\nonumber 
    \\ 
    & + \lambda_{15} (\Phi_2^\dagger \cdot \Phi_4\cdot\Phi_4^\dagger \cdot \Phi_2) + \left(\lambda_{16}(\Phi_2^{\dagger}\cdot\Phi_1)(\Phi_2^{\dagger}\cdot\Phi_2)+ h.c.\right)
 \nonumber \\ &    
 +\left(\lambda_{17}(\Phi_2^{\dagger}\cdot\Phi_1)(\Phi_1^{\dagger}\cdot\Phi_1)+ h.c.\right) 
   +\left(\lambda_{18}(\Phi_2^{\dagger}\cdot\Phi_1)(\Phi_2^{\dagger}\cdot\Phi_1)+ h.c.\right)
   \nonumber
   \\
   &
     + \lambda_{19} (\text{Tr}[\Phi_4^\dagger \cdot \Phi_4])^2 + \lambda_{20} \text{Tr}[(\Phi_4^\dagger \cdot \Phi_4\cdot\Phi_4^\dagger \cdot \Phi_4) ] +  \lambda_{21} (\Phi_1^\dagger \cdot \Phi_1) \text{Tr}[\Phi_4^\dagger \cdot \Phi_4] \nonumber 
    \\ & 
     + \left(\lambda_{22} (\Phi_1^\dagger \cdot \Phi_2) \text{Tr}[\Phi_4^\dagger \cdot \Phi_4]+ h.c.\right) + \lambda_{23} (\Phi_2^\dagger \cdot \Phi_2) \text{Tr}[\Phi_4^\dagger \cdot \Phi_4]  
     +\lambda_{24} (\Phi_3^\dagger \cdot \Phi_3) \text{Tr}[\Phi_4^\dagger \cdot \Phi_4] \nonumber
    \\ & 
    +\left(\lambda_{25}\epsilon_{ijl}  
 \Phi_{4ik}\Phi_{2k}^{*}\Phi_{1j}\Phi_{3l}
+\lambda_{26} \epsilon_{ijl}
\Phi_{4ik}\Phi_{2k}^{*}\Phi_{2j}\Phi_{3l}
+\lambda_{27} \epsilon_{ijl}
\Phi_{4ik}\Phi_{1k}^{*}\Phi_{1j}\Phi_{3l}+ h.c.\right)\nonumber \\
&+\left(\lambda_{28} \epsilon_{ijl}  
 \Phi_{4ik}\Phi_{1k}^{*}\Phi_{3j}\Phi_{2l}+ h.c.\right)
 + \left(\lambda_{29}
\epsilon_{pk'm}
\Phi_{3}^{k'}\Phi_{4}^{mn}
\epsilon_{nk''m'}
\Phi_{3}^{k''}\Phi_{4}^{m'p}+h.c  \right)\nonumber\\
&-\left( \frac{M_1}{\sqrt{2}}\Phi_1^T \Phi_4^\dagger \Phi_1 + \frac{M_1'}{\sqrt{2}}\Phi_1^T \Phi_4^\dagger \Phi_2 +\frac{M_2}{\sqrt{2}}\Phi_2^T \Phi_4^\dagger \Phi_2 + \frac{f}{\sqrt{2}}\epsilon_{ijk}\Phi_1^i \Phi_3^j \Phi_2^k + h.c. \right),
\end{align}
where $M's$ and $f$ have units of mass.\\

To the best of our knowledge, this is the first time that the scalar potential with three triplets and one sextet is presented in its most general form, i.e, without introducing any additional symmetry beyond that dictated by the gauge structure $SU(3)_c \otimes SU(3)_L \otimes U(1)_X$. In order to determine all the possible invariant terms, we have followed two complementary approaches: first, by explicitly constructing each term by hand, and second, by employing the software \texttt{Sym2Int}~\cite{Fonseca:2017lem,Fonseca:2019yya} which systematically generates all tensor products containing a singlet, ensuring automatic consistency with the assigned quantum numbers of the triplets and sextet.  
The explicit construction of the terms of the potential in Eq.~\eqref{eq:Potencial} proportional to $\lambda_{25}$ through $\lambda_{29}$ are presented in Appendix~\ref{sec:su3invariants}.

\subsection{The Yukawa Lagrangian}

With the aim of constructing a realistic model, we must ensure that it can generate the correct known masses for the SM fermion fields and the exotic fields, including the neutral fermions. The Yukawa Lagrangian, as built for the model defined by the fields in the fermionic sector together with the scalar fields introduced in the scalar sector, as allowed by the 3-3-1 symmetries, is given by,

\begin{align}\label{eq:yukawa}
\mathcal{L}_Y \nonumber &=\sum_{\alpha=1}^2Q^3_L\Phi_\alpha C\left( h_\alpha^U U_L^c+\sum_{a=1}^3h_{a\alpha}^u u_L^{ac} \right) + \sum_{i=1}^2Q^i_L\Phi_3^* C\left( h_i^{U\prime} U_L^c+\sum_{a=1}^3h_{i a}^{u\prime} u_L^{ac} \right)\\  &+ \sum_{\alpha=1}^2 \sum_{i=1}^2Q_L^i\Phi_\alpha^*C\left(\sum_{a=1}^3 h_{ia\alpha}^d d_L^{ac}+ \sum_{j=1}^2 h_{ij\alpha}^D D_L^{jc}\right)+Q^3_L\Phi_3 C\left(\sum_{i=1}^2h_i^D D^{ic}_L + \sum_{a=1}^3h_a^d d_L^{ac}\right)\\ \nonumber &+ 
\sum_{i=1}^3 \sum_{j=1}^3 h_{ij}^l L_{iL}\Phi_3 C l_{jL}^c + \sum_{i=1}^3 \sum_{j=1}^3 h_{ij}^{l3} L_{iL} L_{jL}\Phi_3^* + \sum_{i=1}^3 \sum_{j=1}^3 h_{ij}^{l4} {L}_{iL}\Phi_4L_{jL}+H.c.,
\end{align}

\noindent where $C$ is the charge conjugation operator. From the first two rows, the masses of both up-type and down-type quarks are obtained, as well as all observables related to flavor physics in this sector. The first term in the third line generates the masses of the charged leptons of the model, while the second term in the third line generate Dirac-type masses for two of the three neutrinos in the model, although mediated by very small Yukawa couplings in this case ~\cite{Ky:2005yq}. In order to additionally include a Majorana mass term, it is necessary to include a scalar sextet, as given in the last term of the third line. In this last case, with both, Dirac and Majorana mass terms included, not only the model is more general but also the Yukawa couplings do not need to be that small.


\section{The Scalar Fields Mass Matrices}
\label{sec:analisis-analitico}

The 3-3-1 model exhibits a rich structure in its scalar sector. To explore all this features in full, we employ the computational capabilities of \texttt{SARAH}, an open-source package operating within the \texttt{MATHEMATICA} environment~\cite{Staub:2013tta}. This tool facilitates the systematic study of particle physics models. In this framework, all the main features of the model: the multiplets, the Lagrangian, the gauge symmetries and their spontaneous breaking patterns, the Higgs fields acquiring vacuum expectation values (VEV), as well as the CP-even and CP-odd components of the scalar sector and the mass matrix of neutral scalars that remain without a VEV, are included as input. Furthermore, the mass eigenstates of the fermionic sector must be defined.\\

With the implementation of the model in \texttt{SARAH}, the mass matrices related to the scalar sector were obtained in a closed form depending on the free parameters of the model, namely the VEVs and the coupling constants. In this form, we were able to identify the mass of the already known fields as reported in the scientific literature, such as the SM Higgs boson. This will experimentally exclude disfavored values in an ample parameter space, when searching for viable mass ranges for the charged and neutral Higgs bosons that could account for some of the current anomalies in the scalar sector of high-energy physics.\\

Previously, in Ref.~\cite{Tapia:2021uzh}, we implemented  in \texttt{SARAH} a 3-3-1 model with right-handed neutrinos, without exotic electric charges, reproducing successfully analytical results published in the literature, in the case of a scalar sector with only three scalar triplets. This implementation has served as the basis for the present work, in which we have extended the model by incorporating an additional scalar sextet.\\

In this section we present the mass matrices of the charged and neutral scalars of the model, while at the same time performing a counting of the number of Goldstone bosons, as one of the crosschecks for a theoretical consistent model. Given the extension of the full scalar potential in Eq.~(\ref{eq:Potencial}), and without losing generality, we neglect the following coupling constants $\lambda_6, \lambda_{11}, \lambda_{14}, \lambda_{16}, \lambda_{17}, \lambda_{18}, \lambda_{22}, \lambda_{25}, \lambda_{26}, \lambda_{27},\lambda_{28}, \lambda_{29} \text{ and } M_1^{\prime}$. In this limit, the general potential in Eq.~(\ref{eq:Potencial}) reduces to the one presented in Ref.~\cite{Pinheiro:2022obu}. Notice, however, that the trilinear term involving the scalar triplets, with coupling $f$ in Eq.~(\ref{eq:Potencial}) will be present in our future analyses. Several studies in the literature have found arguments to get rid of this term, see for instance Ref.~\cite{Giraldo:2011gd}. Below, we present the analytical mass matrices obtained by executing the \texttt{SARAH} package with the simplest version of the full scalar potential just described above.\\

To obtain the mass matrices, the corresponding scalar fields should be properly shifted around their VEVs:
\begin{equation}
\phi_{i}^{0} \;=\; \tfrac{1}{\sqrt{2}}\big( v_i + h_i + i A_i \big), 
\end{equation}
where $\phi_{i}^{0}$ takes values $\{{\phi}_{1}^0, {\phi}_{2}^{\prime 0}, {\phi}_{3}^{0}, {\phi}_{4}^{\prime 0},{\phi}_{4}^{\prime\prime 0}\}$, $v_i=\{V_1, V_2, v_3, W_2, W_3\}$, $h_i=\{h_4, h_5, h_3, h_7, h_8\}$ and $A_i=\{A_4, A_5, A_3, A_7, A_8\}$. The Type-I seesaw mechanism can be implemented in the model by assuming the $\phi_4^0$ component of the sextet to be inert, i.e., $W_1 = 0$, so that the VEV of $\Phi_4$ takes the form~\cite{Ky:2005yq, Diaz:2003dk}:
\begin{equation}\label{eq:VEVphi4}
 \langle \Phi_4 \rangle=\begin{pmatrix} 0 & 0 & W_2/\sqrt{2} \\ 0 & 0 & 0 \\ W_2 /\sqrt{2} & 0 & W_3 \end{pmatrix}
\end{equation}
It is also necessary to obtain the set of tadpole equations that ensure the reduced potential develops a minimum; these equations are:

\begin{align}\label{eq:Tadpoles}
&\mu_{\phi_3}^2 +  \frac{1}{2}\lambda_{5} V_{1}^{2} +\frac{1}{2}\lambda_{7} V_{2}^{2} + \frac{1}{2}\lambda_{3} v_{3}^{2}  +  \lambda_{24} \Big(W_{2}^{2} + W_{3}^{2}\Big)  - \frac{f V_1 V_2 }{2 \sqrt{2} v_{3}}=0\nonumber\\ 
&\mu_{\phi_1}^2 + \lambda_{1} V_{1}^{2} + \frac{1}{2}\lambda_{4} V_{2}^{2} + \frac{1}{2} \lambda_{5} v_{3}^{2} +\frac{1}{2} \lambda_{21}\Big( W_{2}^{2} +  W_{3}^{2} \Big) + \frac{1}{4}\lambda_{13}W_{2}^{2}  - \frac{f V_2 v_{3}}{2\sqrt{2} V_1} =0\nonumber\\ 
&\mu_{\phi_2}^2 + \frac{1}{2}\lambda_{4} V_{1}^{2} + \lambda_{2} V_{2}^{2}  + \frac{1}{2}\lambda_{7} v_{3}^{2} + \frac{1}{2}\lambda_{15}\Big(\frac{1}{2}W_{2}^{2} + W_{3}^{2}\Big) + \frac{1}{2}\lambda_{23}\Big(W_{2}^{2} + W_{3}^{2}\Big) - \frac{f V_1 v_{3}}{2\sqrt{2} V_2} - M_{2} W_{3}=0\nonumber\\  
&\mu_{\phi_4}^2 + \frac{1}{4}\lambda_{13}  V_{1}^{2}+ \frac{1}{2}\lambda_{21} V_{1}^{2} + \frac{1}{4}\lambda_{15}V_{2}^{2} + \frac{1}{2}\lambda_{23}V_{2}^{2}  + \frac{1}{2}\lambda_{24} v_{3}^{2} + \lambda_{19} \Big( W_{2}^{2} + W_{3}^{2} \Big) + \lambda_{20} \Big(\frac{1}{2} W_{2}^{2} + W_{3}^{2} \Big) =0\nonumber\\ 
& \mu_{\phi_4}^2 +  \frac{1}{2}\lambda_{21} V_{1}^{2}  + \frac{1}{2}\lambda_{15}V_{2}^{2} + \frac{1}{2}\lambda_{23}V_{2}^{2}  + \frac{1}{2}\lambda_{24} v_{3}^{2} + \lambda_{19} \Big( W_{2}^{2} + W_{3}^{2} \Big) + \lambda_{20} \Big( W_{2}^{2} + W_{3}^{2}\Big)   - \frac{M_{2} V_{2}^{2}}{2W_{3}} \Big)=0
\end{align}

Using the values of $\mu_{\phi_i}^2$ obtained from equations above, the mass matrices can now be calculated. The authors in  Ref.~\cite{Pinheiro:2022obu} have considered the case where $f$ and $M_1=M_2$ are around the GeV-scale and obeying a hierarchy $W_2, W_3 \gg M_1, M_2$ that was consistent with their implementation of the Type-II seesaw. In our case we have implemented the Type-I seesaw and therefore different hierarchy for the VEVs is considered. As it will be detailed later on, we have also found $f$ and $M_2$ with values at the GeV-scale as a result of allowing this two parameters to vary in an ample range covering several decades. We have also found a spectrum of scalars analytically (with the aid of \texttt{SARAH}), which at the same time are consistent with the Goldstone theorem, comprising a total of eight Goldstone bosons. 

We begin with the mass matrices of the charged Higgs bosons, taking into account the minimum conditions given in Eq.~(\ref{eq:Tadpoles}). This mass matrix is expressed in the basis $\left(\phi_{2}^{-}, \phi_{3}^{\prime +}, \phi_{4}^{\prime +}\right)$: 
\begin{equation} 
m^2_{H_1^\pm} = \left( 
\begin{array}{ccc}
 \frac{f V_1 v_{3}}{2\sqrt{2}V_{2}}+ \frac{1}{2}\lambda_{10} v_{3}^{2}  &  \frac{f V_1}{2\sqrt{2}} + \frac{1}{2}\lambda_{10} V_2 v_{3}  &0\\ 
 \frac{f V_1}{2\sqrt{2}} +  \frac{1}{2}\lambda_{10} V_2 v_{3}  &  \frac{f V_1 V_{2}}{2\sqrt{2}v_{3}}+ \frac{1}{2}\lambda_{10} V_{2}^{2} & 0\\ 
0 & 0  & -\lambda_{15}V_{2}^{2} + \lambda_{12}v_{3}^{2} + \frac{M_2 V_2^2}{2W_{3}}
\end{array} 
\right). 
 \end{equation} 
When computing the eigenvalues of this matrix, one of them is found to be zero, corresponding to two charged Goldstone bosons.\\

The second mass matrix of the charged Higgs is obtained in the  basis $ \left({\phi}_{1}^{{-}}, {\phi}_{3}^{+}, {{\phi}_{4}}^{{+}}\right)$:

\begin{equation} 
m^2_{H_{2}^{\pm}} = \left( 
\begin{array}{ccc}
 \frac{f V_2 v_{3}}{2\sqrt{2}V_{1}}+\frac{1}{2}\lambda_{9} v_{3}^{2}  & \frac{f V_2}{2\sqrt{2}} +  \frac{1}{2}\lambda_{9} V_1 v_{3}  &0\\ 
\frac{f V_2}{2\sqrt{2}} +  \frac{1}{2}\lambda_{9} V_1 v_{3} &  \frac{f V_1 V_{2}}{2\sqrt{2}v_{3}}+ \frac{1}{2}\lambda_{9} V_{1}^{2} & 0\\ 
0 & 0  & -\frac{1}{4}\lambda_{15}V_{2}^{2} + \frac{1}{4}\lambda_{12}v_{3}^{2} 
\end{array} 
\right).
 \end{equation} 
Again, when computing the eigenvalues of this matrix, one of them is found to be zero, corresponding to two charged Goldstone bosons.\\

We now proceed with the neutral scalar sector. The model contains eight neutral fields, however, we assume that three of them are inert (i.e., have no VEV), namely $\phi_4^0$, $\phi_1^{\prime 0}$, and $\phi_2^0$. As a consequence, these fields decouple from the remaining five, and, considering the basis $(\phi_4^0, \phi_1^{\prime 0}, \phi_2^0)$, the following $3 \times 3$ mass matrix for these scalars is obtained:

\begin{equation} 
m^2_{H^{0}} = \left( 
\begin{array}{ccc}
\frac{1}{2}\Big(\frac{M_2V_2^2}{W_3}+\lambda_{13}V_{1}^{2} - \lambda_{15}V_{2}^{2}\Big)  & 0  & 0 \\ 
0  & \frac{1}{4} \Big(2 \lambda_{8}V_2^2  + \frac{\sqrt{2} f V_2 v_{3}}{V_1} \Big) & -\frac{1}{4} \Big(2 \lambda_{8} V_1 V_2  + \sqrt{2} f v_{3} \Big)\\ 
0 &-\frac{1}{4} \Big(2 \lambda_{8} V_1 V_2  + \sqrt{2} f v_{3} \Big) & \frac{1}{4} \Big(2 \lambda_{8}V_1^2  + \frac{\sqrt{2} f V_1 v_{3}}{V_2} \Big)\end{array} 
\right).
 \end{equation} 

When computing the eigenvalues of this matrix, one of them is found to be zero, corresponding to one neutral Goldstone boson.\\

We now consider the CP-odd fields, organized in the basis $\left(A_8, A_7, A_5, A_4, A_3\right)$. After applying the minimization conditions of the potential, the following mass matrix is obtained:

\begin{equation} 
m^2_{A} = \left( 
\begin{array}{ccccc}
 -\frac{1}{4}\lambda_{13}V_1^2+ \frac{1}{4}\lambda_{15}V_2^2& 0 &M_{2} V_2  & 0 & 0\\ 
0 & 0 & 0 & 0 & 0\\ 
M_{2} V_2  & 0 & \frac{fV_1v_3}{2\sqrt{2}V_2}+ M_2W_3 & \frac{f v_{3}} {2\sqrt{2}}  & \frac{f V_{1}} {2\sqrt{2}}\\ 
0 & 0 &  \frac{f v_{3}} {2\sqrt{2}}  &  \frac{fV_2v_3}{2\sqrt{2}V_1} & \frac{f V_{2}} {2\sqrt{2}} \\ 
0 &0 & \frac{f V_{1}} {2\sqrt{2}} & \frac{f V_{2}} {2\sqrt{2}} & \frac{fV_1V_2}{2\sqrt{2}v_3}\end{array} 
\right),
 \end{equation} 
 when computing the eigenvalues of this matrix, two of them are found to be zero, corresponding to two neutral Goldstone bosons.\\

We now turn to the CP-even sector, considering the basis \( \left(h_8, h_7, h_5, h_4, h_3\right) \) and taking into account the minimum condition given in Eq.~\ref{eq:Tadpoles} we obtain:
 
\begin{equation}\label{eq:ScalarMassMatrix}
m^2_{H} = \left( 
\begin{array}{ccccc}
-\frac{1}{4}\lambda_{13}V_1^2 +\frac{1}{4}\lambda_{15}V_2^2 & 0 &-M_2 V_2 & 0 & 0 \\ 
0 & 0 & 0 & 0 & 0 \\ 
-M_2 V_2 & 0 & 2\lambda_2 V_2^2 + \frac{fV_1v_3}{2\sqrt{2}V_2}  & \lambda_4 V_1 V_2 - \frac{fv_3}{2\sqrt{2}}   &  \lambda_7 V_2 v_3 - \frac{f V_1}{2\sqrt{2}} \\ 
0  & 0 &  \lambda_4 V_1 V_2 - \frac{fv_3}{2\sqrt{2}}  &  2\lambda_1 V_1^2 + \frac{fV_2v_3}{2\sqrt{2}V_1} &  \lambda_5 V_1 v_3 - \frac{f V_2}{2\sqrt{2}}\\ 
0 & 0 & \lambda_7 V_2 v_3 - \frac{f V_1}{2\sqrt{2}} & \lambda_5 V_1 v_3 - \frac{f V_2}{2\sqrt{2}} & 2\lambda_3 v_3^2 + \frac{fV_1V_2}{2\sqrt{2}v_3}\end{array} 
\right).
 \end{equation} 
When computing the eigenvalues of this matrix, one of them is found to be zero, corresponding to one neutral Goldstone bosons.\\
  
By counting the Goldstone bosons, we obtain a total of 4 charged and 4 neutral states, yielding eight Goldstone bosons in total. Four neutral are used to provide with masses the four electrically neutral gauge bosons ($Z^0$, $Z^{'0}$, $K^0$ and $\bar{K}^0$), and the four charged ones are used to provide with masses to $W^{\pm}$ and to $K^{\pm}$. This shows the theoretical consistency of our results.\\

Finally, in this model, there is only one doubly charged scalar, $\phi_4^{++}$, whose mass at tree level is given by:

\begin{equation}
    m^2_{\phi_4^{++}}= \frac{1}{2}\Big( \lambda_{15} V_2^2 - \lambda_{12} v_3^2 + 2\lambda_{20} (W_2^2 + W_3^2) - \frac{M_2 V_2^2}{W_3} \Big).
\end{equation}\\

Searches for a  doubly charged scalar have been pursued at the LHC in Refs.~\cite{ATLAS:2021jol, ATLAS:2022pbd}, and have also been the subject of the study in a 3-3-1 model in Ref.~\cite{CiezaMontalvo:2006zt}.

\section{Probability Distributions of Scalar Masses\label{sec:pdf}}
Before presenting the main results of this section. It is worth to comment on the status of the Higgs-like ressonances.\\

In Ref.~\cite{Crivellin:2023zui}, several experimental anomalies in processes with diphotons in the final state are discussed. Ordered by their reported significance, these are:
\begin{itemize}
  \item \textbf{Excess near 152~GeV (diphoton--associated production).}  
  A combined/global significance of about $3.9\sigma$ is quoted when
  different channels are combined under a simplified model
  $pp \to H \to SS^\ast$.  
  \emph{Comment:} this anomaly is mainly observed in $\gamma\gamma$
  final states with missing energy or associated production~\cite{Banik:2024ftv}.  
  \item \textbf{Excess near 95~GeV (diphoton resonance).}  
  A combined/global significance of about $3.8\sigma$ is reported after taking into account supporting hints in $\tau\tau$, $ZH$ ($H\to b\bar b$), and $WW$ channels.  \emph{Comment:} this is the most long–standing low-mass diphoton hint, consistently appearing in CMS data and supported by LEP and ATLAS
 observations~\cite{Ashanujjaman:2023etj}.  

  \item \textbf{Di-Higgs--like structure at $\sim 650$~GeV ($b\bar b + \gamma\gamma$).}  
  A local significance of about $3.8\sigma$ is reported, which reduces to a global significance of $2.8\sigma$ after look–elsewhere effects.  
  \emph{Comment:} the invariant mass of the $b\bar b$ pair in these events is compatible with $\sim 95$~GeV, linking it to the previous anomaly~\cite{Crivellin:2023zui}. 

  \item \textbf{Hints near $\sim 680$~GeV in $\gamma\gamma$ and $ZZ$ channels.}  
  Each of these excesses reaches about $3\sigma$ local significance.  
  \emph{Comment:} these are weaker hints compared to the 95~GeV and
  152~GeV anomalies~\cite{Crivellin:2023zui}.
\end{itemize}

It is important to emphasize that LEP does not exclude a 95 GeV scalar; on the contrary, the LEP data show a marginal excess consistent with such an anomaly. 
The combined LEP Higgs searches~\cite{LEPWorkingGroupforHiggsbosonsearches:2003ing} established the
$95\%$ C.L. lower bound of $114.4$~GeV on a Standard Model--like Higgs boson.
Interestingly, the analysis also reported a small local excess near
$98$~GeV, which has since motivated interpretations in terms of
non--standard scalar states.
This result must be compared with the CMS low-mass diphoton analysis~\cite{CMS:2018cyk} studied the range $70$--$110$~GeV using the full Run~2 dataset at $\sqrt{s}=13$~TeV (together with earlier $8$~TeV data), and reported a local excess at about
$95$~GeV with a significance close to $2.9\sigma$.
In parallel, the ATLAS diphoton resonance search with the full Run~2
statistics~\cite{ATLAS:2023jzc} found a mild excess in the same mass
region, around $95$~GeV with local significance $\sim 1.7\sigma$, 
consistent with the CMS observation but not sufficient for evidence.
Previous works have analyzed these features~\cite{Biekotter:2022abc,Heinemeyer:2021msz}. \\
Now we move to the main results of this section.
The neutral scalar mass matrix squared, computed in \texttt{SARAH}, appears in Eq~(\ref{eq:ScalarMassMatrix}). Physical masses can then be obtained by diagonalizing this matrix, after assuming phenomenologically viable values for the VEVs,

\begin{equation}\label{eq:vevVals}
    \begin{split}
        v_3&=100\,\text{GeV} \\
        V_1&=\sqrt{v_{SM}^2-v_3^2}\approx 224.76\,\text{GeV}\\
        V_2&=6\times 10^3\,\text{GeV}\,.
    \end{split}
\end{equation}

Without loss of generality, it can be assumed that the value of the couplings in the potential in Eq.~(\ref{eq:Potencial}) that appear in Eq~(\ref{eq:ScalarMassMatrix}) are real. Therefore, in this case, the number of parameters in Eq.~(\ref{eq:ScalarMassMatrix}) are, eight dimensionless parameters $\lambda$'s, and two dimension-full parameters ($M_2$ and $f$), for a total of ten free parameters. Without external constraints, the only requirement at this point is that one of the mass eigenvalues must correspond to the SM Higgs boson mass, which has a central value of $125.0$~GeV \cite{ParticleDataGroup:2024cfk}.\\

\begin{figure}[h!]
\centering 
\includegraphics[width=0.49\textwidth]{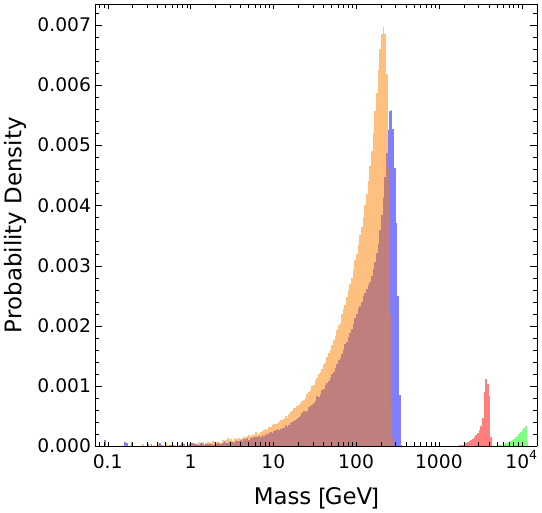}
\caption{\label{fig:scalarmasses} Mass distributions for the four non-zero mass eigenvalues of Eq~(\ref{eq:ScalarMassMatrix}).}
\end{figure}

Given the large number of free parameters involved, we have, initially, performed a random scan where the $\lambda$'s were allowed to vary uniformly within the interval $[-2,2]$, as requiered by perturbativity. The dimensionfull parameters, $M_2$ and $f$, were allowed to vary in the intervals $[10^3,10^{15}]$~GeV and $[-10^{10},10^{10}]$~GeV, respectively. Then, we selected physical values for the ten couplings requiring only that all the four non-zero eigenvalues of Eq~(\ref{eq:ScalarMassMatrix}) be positive. The resulted distributions for the non-zero eigenvalues are shown in Fig.~\ref{fig:scalarmasses}.\\

\begin{figure}[h!]
\centering
\includegraphics[width=0.49\textwidth, height=8cm]{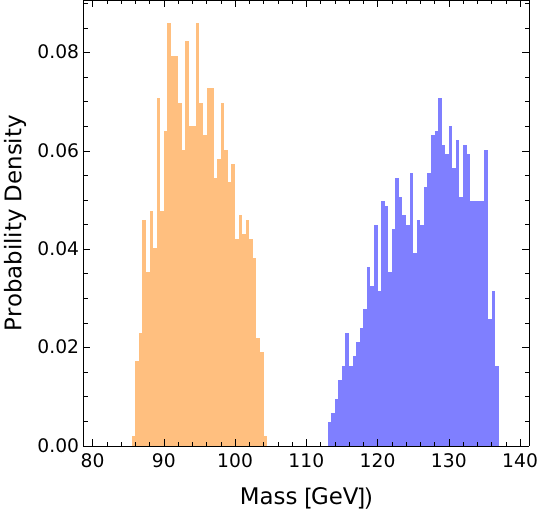}
\includegraphics[width=0.49\textwidth, height=8cm]{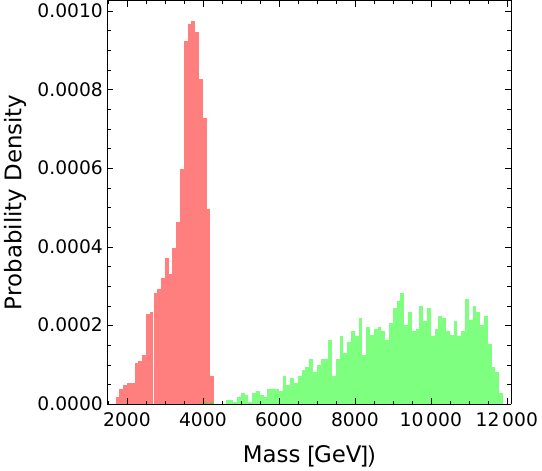}
\caption{\label{fig:HigssMass} Mass distributions for the four non-zero mass eigenvalues, in the same color code as in Fig.~\ref{fig:scalarmasses} after the SM Higgs mass is assigned to one of the two low mass distributions while the other one was assigned to an scalar mass of $95$~GeV (see the text for more details). In the left panel, two low-mass distributions, one corresponding to a mass of $95$~GeV and the other to the SM Higgs mas of $125$~GeV. In the right panel, the remaining two mass distributions, where the mean values of the mass distributions are $3473$~GeV and $9186$~GeV, respectively.}
\end{figure}

From Fig.~\ref{fig:scalarmasses}, one can see that the Higgs mass can be assigned to any of the two low mass distributions, which overlap. After, this assignment, there is still room for another neutral scalar with mass of the same order as the SM Higgs mass. Using the physical ranges for the ten couplings, we performed a second scan selecting physical values for the couplings that were also compatible with the SM Higgs mass within an ample deviation of 5\%, and at the same time, compatible with a light neutral scalar with a mass of $95$~GeV within a deviation of 5\%. The distributions in this case are shown in Fig.~\ref{fig:HigssMass}.\\
 
After the SM Higgs and the $95$~GeV mass assignements, as shown in Fig.~\ref{fig:HigssMass}. (second analysis), the following values for the expectation plus/minus the squared root of the variance from the mass distributions are: $(95 \pm 5)$~GeV, $(127 \pm 6)$~GeV, $(3473 \pm 518)$~GeV, and $(9186 \pm 1605)$~GeV. Given that the mass distributions are not gaussian, we also give, as a reference, a point in the ten-dimensional parameter space in the third column in Table~\ref{tab:10paramsRange} as well as the full range for each parameter in the second analysis (second column in the same table), from which the results in Fig.~\ref{fig:HigssMass} were obtained. Notice that the range of some of the couplings are constrained after the mass assignments for the SM Higss and the $95$~GeV resonance, in particular the dimensionfull couplings $f$ and $M_2$. It is worth noting that, unlike Ref.~\cite{Palcu:2019nld}, we have not assumed the $f$ coupling to be proportional to any of the VEVs considered in Eq.~(\ref{eq:vevVals}). Instead, we have assumed this parameter to be free and we have found it is smaller than any of the VEVs considered, as can be seen from Table~\ref{tab:10paramsRange}.\\

\begin{table}[h!]
\centering
\begin{tabular}{c|c|c} \hline
Parameter & Range & Reference point \\ \hline
   $\lambda_{1}$ & $[0.12,1.12]$ & 0.278 \\
   $\lambda_{2}$ & $[0.21,1.83]$ & 0.735\\
   $\lambda_{3}$ & $[0.03,0.44]$ & 0.441\\
   $\lambda_{4}$ & $[-2.0,-0.43]$ & -0.713 \\
   $\lambda_{5}$ & $[1.02,2.0]$ & 1.476\\
   $\lambda_{7}$ & $[-2.0,-0.86]$ & -1.932\\
   $\lambda_{13}$ &  $[-1.99,0.84]$ & -0.096 \\
   $\lambda_{15}$ & $[1.42,2.0]$ & 1.580\\
   $f$~[GeV] & $[1.33,10.50]$ & 5.949 \\
   $M_2$~[GeV] &  $[-1691.24,5049.05]$ & 1230.5\\ \hline
\end{tabular}
\caption{\label{tab:10paramsRange} Ranges of the couplings that enter in Eq.~(\ref{eq:ScalarMassMatrix}) obtained in the second analysis, reported in the second column. In the third column, a reference point in the 10-dimensional parameter space is also provided. This point will give fixed values for the neutral scalar masses compatible with the desired phenomenology, see text for details.}
\end{table}

The parameter values in the third column in table~\ref{tab:10paramsRange} will produce the following values for the neutral scalar masses: $95$~GeV, $125$~GeV, $3587$~GeV, and $7371$~GeV. This confirms, the model have enough freedom not only to accomodate the SM Higgs mass but also a lighter neutral scalar with a mass of $95$~GeV. From this analysis, also two other neutral scalars with larger masses of $3587$~GeV, and $7371$~GeV are obtained.


\section{Neutrino Yukawa Couplings in the Type-I Seesaw\label{sec:seesaw}}

As shown in the works of Ref.~\cite{Benavides:2009cn}, all charged fermions, and even the gauge bosons of the model, acquire their masses after the spontaneous breaking of the symmetry through the interaction with the three scalar triplets. However, all neutral fermions remain massless at tree level, requiring the inclusion of additional scalar fields in order to generate their masses radiatively. In the present case, we employ the scalar sextet to generate the masses of the three light neutrinos via the Type-I seesaw mechanism.

Once the spontaneous symmetry breaking has occurred for all components of the scalar sector, the neutrino mass matrix in the interaction basis takes the form:

\begin{equation}\label{eq:MnMatrix}
    h_{ab}^{l4}(\bar{\nu}^{a}_L,\bar{N}^{aC}_L) \begin{pmatrix} 0 & W_2/\sqrt{2} \\ W_2/\sqrt{2} & W_3 \end{pmatrix} 
    \begin{pmatrix} \nu_L^{bC} \\ N_L^{bC} \end{pmatrix}
\end{equation}

or alternatively:

\begin{equation}\label{eq:numasss}
    \frac{1}{2}(\bar{\nu}^{a}_L,\bar{N}^{aC}_L) \begin{pmatrix}0 & m_D \\ m_D^T & m_R \end{pmatrix} 
    \begin{pmatrix} \nu_L^{bC} \\ N_L^{bC} \end{pmatrix},
\end{equation}

where $m_D$ is used to describe Dirac-type masses, and $m_R$ denotes the Majorana mass term. The mass matrix exhibits the structure of a Type-I seesaw, which can be diagonalized to obtain the physical neutrino masses.\\

The neutrino mass matrix in the Type-I seesaw in Eq.~(\ref{eq:numasss}) was written in the standard form:
\begin{equation}\label{eq:seesaw}
M_\nu=
    \begin{pmatrix} 
    0 & m_D \\ 
    m_D^T & m_R 
    \end{pmatrix}.
\end{equation}

In the model considered in this work, with $W_1 \ne 0$, an additional mass term would be present instead of the zero block in the $(1,1)$-element, and therefore the Type-II seesaw can also be realized in the model~\cite{Pinheiro:2022obu}, althouth with a different VEV hierarchy between the $W$'s. Differently, in this work, we have assumed the hierarchy $W_1 \ll W_2 \ll W_3$ such that neutrinos get its mass via the Type-I seesaw (see Eq.~(\ref{eq:MnMatrix})), and for the numerical results, we have fixed $W_1=0$ (as noted before in Eq.~(\ref{eq:VEVphi4})). After a block diagonalization of Eq.~(\ref{eq:seesaw}) (using a perturbative expansion), the light neutrino mass, at leading order, is given by~\cite{Schechter:1981cv}:

\begin{equation}\label{eq:lightnumass}
    m_\nu \approx -m_D\,m_R^{-1} m_D^T\,,
\end{equation}
while the heavy neutrino mass is $m_H \sim \text{diag}(m_R)\equiv \hat{m}_R$, since the assumed hierarchy implies $m_D \ll m_R$.
It is usually considered that to be compatible with the observed active neutrino mass of the order of $\sim 1\,\text{eV}$, and {\it with Yukawa couplings of order one} then $m_D \lesssim 174\,\text{GeV}$, which implies $m_R \gtrsim 3\times 10^{13} \text{GeV}$. Therefore, in the Type-I seesaw, the smallness of the neutrino mass results from the existence of a mass-scale (for $m_R$) close the GUT scale \cite{Gell-Mann:1979vob,Yanagida:1979as, Mohapatra:1979ia}. However, this scale can be moved to a lower one at the expense of the Yukawa couplings, which needs to be smaller than one such that the light neutrino mass scale of $~1\,\text{eV}$ is still accomplished. We will be back to this point of rescaling the heavy neutrino mass scale later on.\\

It is possible to write the Dirac mass $m_D$ in terms of the observed lepton mixing matrix $U$, the light neutrino masses, and the scale of the heavy neutrinos, what is known as the Casas-Ibarra parametrization~\cite{Casas:2001sr}. To achieve this, the Lagrangian is written on a basis where the Majorana mass is diagonal $m_R=\hat{m}_R$, then starting from Eq.~(\ref{eq:lightnumass}), and omitting the negative global sign,
\begin{equation}\label{eq:Casas-Ibarra}
    m_D=U\sqrt{\hat{m}_\nu}\,R^T\,\sqrt{\hat{m}_R}\,,
\end{equation}

where $\hat{m}_\nu=U^T m_\nu U$ are the light neutrino masses, and $R$ is an orthogonal matrix, i.e, $R^T=R^{-1}$. \\

After the implementation of the model in \texttt{SARAH}, the $(2,2)$ block corresponding to $m_R$, is a general symmetric matrix (as it should be for a Majorana mass) but it is not diagonal, and therefore the Casas-Ibarra parametrization in Eq.~(\ref{eq:Casas-Ibarra}) can not be applied directly. There are then two options to proceed, to set a specific basis in  \texttt{SARAH} such that the Majorana mass is diagonal or to use a modified expression of Eq.~(\ref{eq:MnMatrix}) accounting for non-diagonal $m_R$. We have followed the latter path, assuming that there is an orthogonal matrix that diagonalizes $m_R$, such that $\hat{m}_R= U_R^T m_R U_R$ from where a modified Casas-Ibarra expression is obtained~\cite{Casas:2001sr},
\begin{equation}\label{eq:Casas-Ibarra_v2}
    m_D=U\sqrt{\hat{m}_\nu}\,R^T\,\sqrt{\hat{m}_R}\,U_R^T\,.
\end{equation}
Although it might look simpler, it is not trivial to find a matrix $U_R$, in a closed form, that diagonalizes a general $3\times3$ real symmetric matrix. We have used a procedure, involving {\it diagonalization invariants}, described in the Appendix~\ref{sec:appendixDiagonal}. This method allows to find some of the $U_R$ elements in terms of the diagonal values $\hat{m}_R$ (which are proportional to the heavy neutrino masses), and therefore, some of the $U_R$ elements remain unknown. It is worth mentioning that the method gives several solutions. For the numerical analysis, we have chosen a solution for the $U_R$ elements where four out of the six $m_R$ elements were written in terms of only two of them. Finally, requiring the $U_R$ elements be finite, the eigenvalues $\hat{m}_R$ must safisfy $(\hat{m}_R)_1\ne(\hat{m}_R)_2\ne(\hat{m}_R)_3$, i.e, can not be degenerated. Therefore, in the case where the heavy neutrino masses were degenerated $U_R \sim I$ in Eq.~(\ref{eq:Casas-Ibarra_v2}) and the usual form of the Casas-Ibarra parametrization is recovered.\\

In the numerical analysis, we have allowed most of the parameters to vary randomly assuming a uniform distribution for each of them. From Eq.~(\ref{eq:Casas-Ibarra_v2}) we have five oscillation parameters (after fixing the Dirac CP-phase to zero), plus the absolute neutrino mass (which has an upper bound from direct mass searches $m_0<0.45$~eV at 90\% of C.L, as the case of KATRIN experiment~\cite{KATRIN:2024cdt}), plus three free parameters in $R$ (assumed real), plus two free parameters in $U_R$, and a global scale factor that multiply the heavy neutrino masses (since the heavy neutrino masses were fixed), for a total of twelve parameters. The oscillation parameters were allowed to vary within their one-sigma range from~\cite{deSalas:2020pgw}, the heavy neutrino masses were fixed to the values $\hat{m}_R=(1, 2, 10)\,\text{TeV}$ multiplied for a free scale $\alpha=1\to 10^{10}\,\text{GeV}$. However, light neutrino masses are not sensitive to this scale, as we explain below.\\

Let us comment on the issue of size of the Yukawa couplings, after a rescaling of the heavy neutrino masses. In the Type-I seesaw, the light neutrino mass is approximately given by Eq.~(\ref{eq:lightnumass}) and this expression does not change under a simultaneous scaling of the Dirac and Majorana mass in the form $m_R \to \alpha \, m_R$ and $m_D \to \sqrt{\alpha}\, m_D$. The structure of $m_R$ in the neutrino mass matrix in Eq.~(\ref{eq:MnMatrix}) is precisely a general symmetric matrix multiplied by the $W_3$ VEV. It is then convenient to calculate the $m_D$ element distributions independent of $W_3$, in the form $(m_D)_i/\sqrt{W_3}$. Since all the elements of $m_D$ are multiplied by the $W_2$ VEV, the Yukawa couplings can be calculated independently of both VEV factors, and the full result is recovered multiplying by the factor $\sqrt{W_3}/W_2$.

After scanning over the parameters, the distributions of the $m_D$ elements were nearly Gaussian, although we did not apply any Gaussianity test. In any case, independent of the underlying distribution, here we report the allowed values for the $m_D$ elements within a one sigma range for 1~d.o.f without assuming any distribution in particular:

\begin{equation}
\frac{m_D}{\sqrt{W_3}}=
 \begin{pmatrix}
     [-0.99, 0.99] & [-0.77, 0.77] & [-1.08, 1.08] \\
     [-0.89, 0.89] & [-0.74, 0.74] & [-0.98, 0.98] \\
     [-0.92, 0.92] & [-0.75, 0.75] & [-1.01, 1.01] \\
 \end{pmatrix}
 \times 10^{-3} \,\sqrt{\text{GeV}}\,,
\end{equation}
where all the elements are compatible with a central value equal to zero and the interval of confidence is symmetric for each element. From the Yukawa lagrangian in Eq.~(\ref{eq:yukawa}), the constraints on $m_D$ can be written as:

\begin{equation}\label{eq:Yukawas}
|h^{\nu \,4}_{ij}| \le \frac{\sqrt{W_3}}{W_2} \times
 \begin{pmatrix}
     0.99 & 0.77 & 1.08 \\
     0.89 & 0.74 & 0.98 \\
     0.92 & 0.75 & 1.01 \\
 \end{pmatrix}
 \times 10^{-3} \,\sqrt{\text{GeV}}\,.
\end{equation}

This result is quite general for the neutrino mass in Eq.~(\ref{eq:MnMatrix}), which has a Type-I seesaw structure. With this structure, the Yukawa couplings in Eq.~(\ref{eq:Yukawas}) can be reported independently of the $W_3$ and $W_2$ scales as long as the hierarchy $W_1 \ll W_2 \ll W_3$ is maintained such that the light neutrino mass is given by the expression in Eq.~(\ref{eq:lightnumass}).

\section{Flavor and Electroweak Constraints\label{sec:ew-fcnc}}

For a Beyond Standard Model scenario with an extended electroweak sector, the neutral current interactions of the fermions are described by the Hamiltonian.
\begin{align}\label{eq:hnc}
 H_{NC}=&  \sum_{i=1}^2 g_i Z_{i\mu}^0
 \sum_f\bar{f}\gamma^{\mu}\left(\epsilon_{iL}(f)P_L+\epsilon_{iR}(f)P_R\right)f,
 \end{align}

where $f$ runs over all the SM fermions in the low energy Neutral Current (NC) 
effective Hamiltonian $H_{NC}$, and $P_L=(1-\gamma_5)/2$ and $P_R=(1+\gamma_5)/2$. For 3-3-1 models, 
the relationship between $g_1$ and $g_2$ is model dependent, but for all the cases we can write
 \begin{align}\label{eq:331hnc}
H_{NC} =&\frac{g}{2 \cos \theta_W}\sum_{i=1}^2  Z_{i\mu}^0
 \sum_f\bar{f}\gamma^{\mu}\left(g_{iV}(f)-g_{iA}(f)\gamma_5\right)f,\,
\end{align}
\noindent
where the chiral couplings $\epsilon_{iL}(f)$ and $\epsilon_{iR}(f)$ are linear combinations of 
the vector $g_{iV}(f)$ and axial $g_{iA}(f)$ charges given by $\epsilon_{iL}(f)=[g_{iV}(f)+g_{iA}(f)]/2$ 
and $\epsilon_{iR}(f)=[g_{iV}(f)-g_{iA}(f)]/2$.

The physical fields in the former expressions are:
\begin{eqnarray*}
 Z_1^\mu&=& \ \ Z^\mu\cos\theta+Z^{\prime\mu}\sin\theta, \\
 Z_2^\mu&=& -Z^\mu\sin\theta+Z^{\prime\mu}\cos\theta,
\end{eqnarray*}
where $Z^\mu$ and $Z^{\prime\mu}$ are the weak basis states such that $Z^{\mu}$ is identified with the neutral 
gauge boson of the SM. At a first approximation we have taken $\theta=0$.

For the numerical calculations, we use the SM values:
$M_W = 80.401$ GeV, $M_Z = 91.188$ GeV,  $\cos\theta_W = M_W/M_Z$,
$\delta= \sqrt{4\cos^2\theta_W-1}$ and $g_1\equiv g/\cos \theta_W = 0.7433$~\cite{ParticleDataGroup:2024cfk}.

\begin{table}[ht]
\centering
\label{tab:amodel}
\begin{tabular}{c c c}
\hline
\hline 
Field & $g_V$ & $g_A$ \\\hline
$\nu_\alpha$ & $(\frac{1}{2}-\sin^2\theta_W)\frac{1}{\delta}$ &  $(\frac{1}{2}-\sin^2\theta_W)\frac{1}{\delta}$ \\
$e_\alpha$ & $-(-\frac{1}{2}+2\sin^2\theta_W)\frac{1}{\delta}$ &  $\frac{1}{2}\frac{1}{\delta}$ \\
$u_i$ & $(-\frac{1}{2}+\frac{4}{3}\sin^2\theta_W)\frac{1}{\delta}$ & $-\frac{1}{2}\frac{1}{\delta}$ \\
$u_3$ & $(\frac{1}{2}+\frac{1}{3}\sin^2\theta_W)\frac{1}{\delta}$ & $-(-\frac{1}{2}+\sin^2\theta_W)\frac{1}{\delta}$ \\
$d_i$ & $-(\frac{1}{2}-\frac{1}{3}\sin^2\theta_W)\frac{1}{\delta}$ & $-(\frac{1}{2}-\sin^2\theta_W)\frac{1}{\delta}$\\
$d_3$ & $-(-\frac{1}{2}+\frac{2}{3}\sin^2\theta_W)\frac{1}{\delta}$ &  $\frac{1}{2}\frac{1}{\delta}$\\
\hline
\end{tabular}
\caption{Vector and axial couplings in the 3-3-1 model with right-handed neutrinos, $\alpha=1,2,3,$ and $i=1,2$}
\end{table}

\begin{figure}[ht]

\centering 
\begin{tabular}{c}
\includegraphics[scale=0.28]{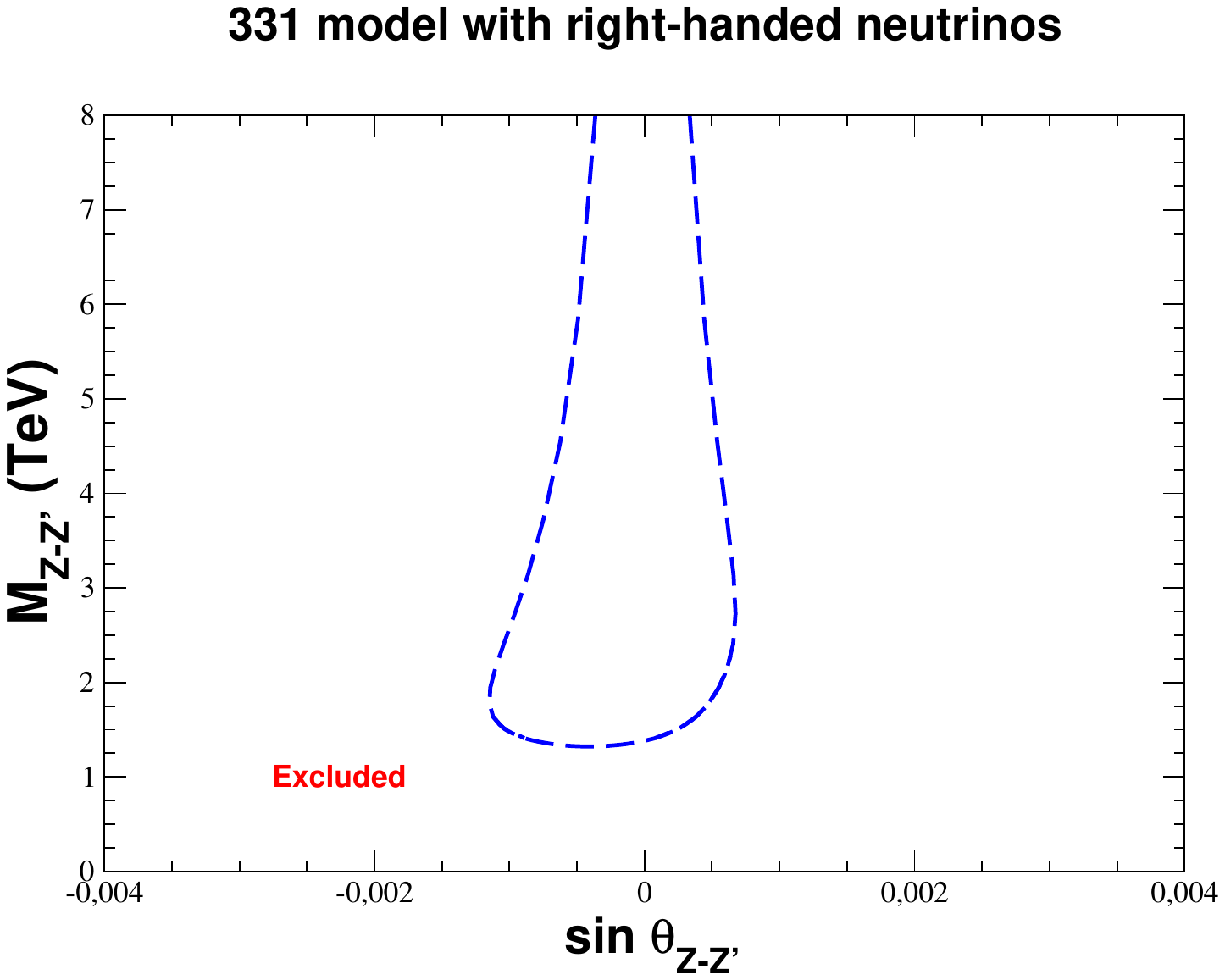}
\end{tabular}
\caption{95\% C.L. contours in $M_{Z'}$ vs. $\sin \theta_{ZZ'}$ for the 3-3-1 model with right-handed neutrinos. See the text for details.}
\label{Contours1}
\end{figure}
From reference~\cite{Erler:2009jh} the $Z-Z'$ mixing angle is constrained to be $\theta_{Z-Z'}< 10^{-3}$ for $E_6$ models. These bounds come from 
Z-pole precision observables,  atomic parity-violating,  deep inelastic neutrino-nucleon scattering, W  and top quark masses, and other less precise low-energy experiments.
We apply this analysis to the model considered in the present work.
The 95 CL allowed region is shown in Figure~\ref{Contours1}.

This model is non-universal in the quark sector, since the $Z'$ charges assigned to the third generation differ from those of the first two. As a result, the diagonalization matrices that rotate from the interaction basis to the mass basis induce tree-level Flavor Changing Neutral Currents (FCNCs), which can in turn provide stringent constraints on the model.
  If  the mass eigenstates $q'$ are related to those of the interaction  basis $q$  by a unitary transformation, i.e., $u'=U^{U}u$ and  $d'=U^{D}d$, the CKM mixing matrix will be given by $V^{\text{CKM}}=U^{U\dagger}U^{D}$. Where $U^{U}$ corresponds to the diagonalization of the up-type quark mass matrix, and it can be written as
\begin{align}\label{eq:rab}
U^{U}=R_1(\beta_3)R_2(\beta_2)R_3(\beta_1)\ ,
\end{align}
where the $R_i$ are block matrices given by
\begin{subequations}
\label{eqC4}
\begin{equation}
R_1(\beta_1)=
\begin{pmatrix}
 \cos \beta_1 & \sin \beta_1  & 0\\
-\sin \beta_1 & \cos \beta_1  & 0\\
      0     &   0        & 1     
\end{pmatrix},
\end{equation}
\begin{equation}
R_2(\beta_2)=
\begin{pmatrix}
 \cos \beta_2    &   0   & \sin \beta_2e^{-i\delta} \\          
      0          &   0   &  0           \\
 -\sin \beta_2 e^{i\delta}  &   0   & \cos \beta_2 
\end{pmatrix},
\end{equation}
\begin{equation}
R_3(\beta_3)=
\begin{pmatrix}
        1              &   0          & 0              \\ 
        0              & \cos \beta_3 & \sin \beta_3   \\
        0              &-\sin \beta_3 & \cos \beta_3   
\end{pmatrix}\ .
\end{equation}
\end{subequations}
Without loss of generality, the diagonalizing operator of the down-type mass matrix is given by $U^D= U^U V^{CKM}$. 
  For $Z'$ couplings of the left-handed down-type quarks, \textit{i.e.,} $d$, $s$ and $b$  we use    $\Delta^{D}_{L}=g_{z'} \left(U^UV^{CKM}\right)^\dagger 
  \epsilon_{L}^{Z'}(d)
  \left(U^UV^{CKM}\right)$.
\begin{figure}[t]
\begin{center}
\centering 
\begin{tabular}{ccc}
\includegraphics[scale=0.3]{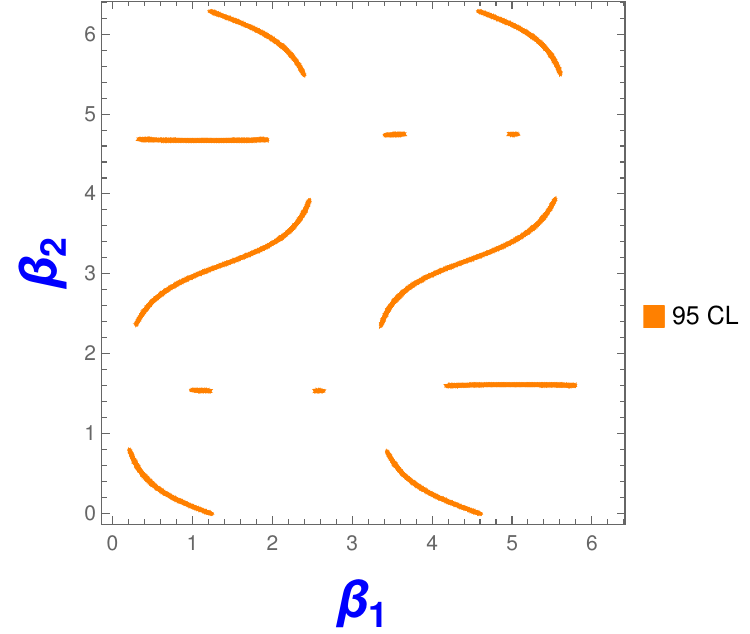} & 
\includegraphics[scale=0.3]{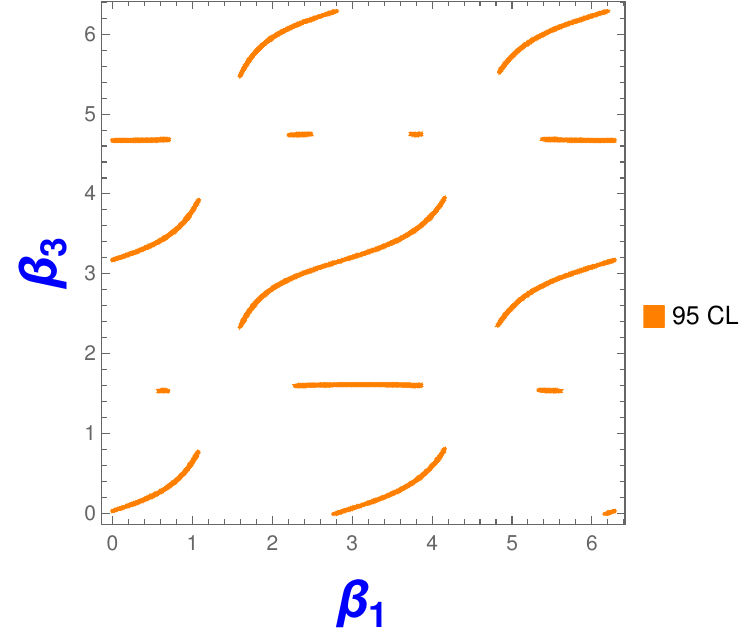} &
\includegraphics[scale=0.3]{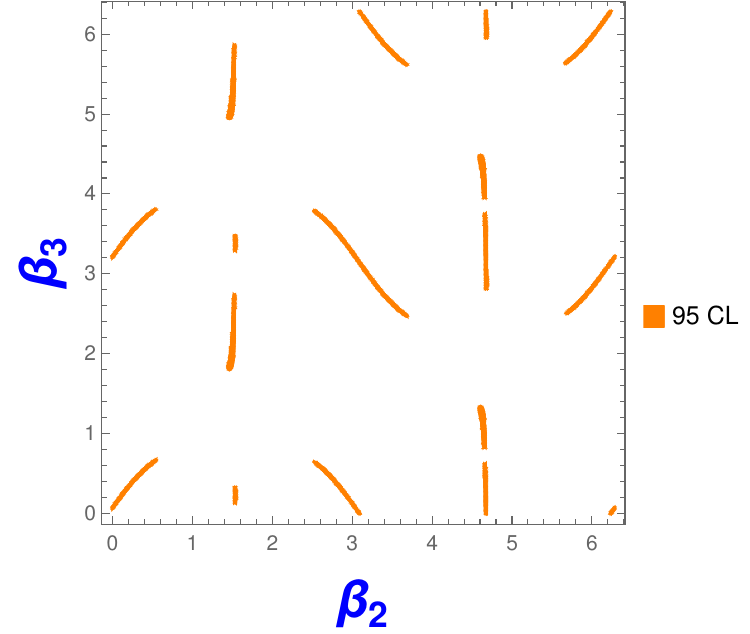}\\
\includegraphics[scale=0.3]{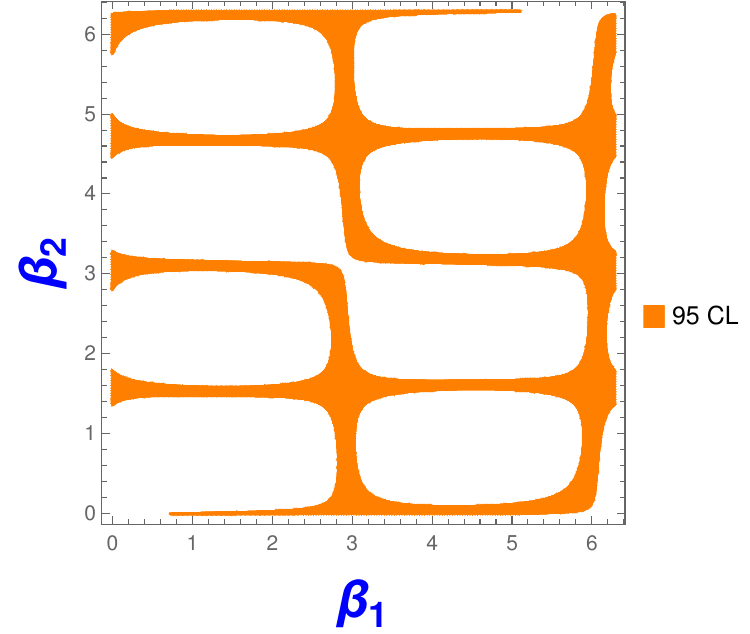} &  \includegraphics[scale=0.3]{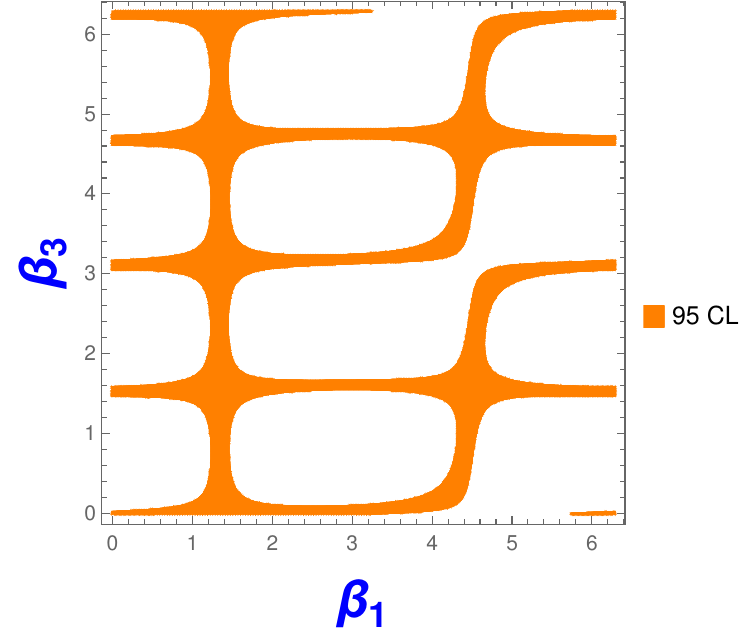} &
\includegraphics[scale=0.3]{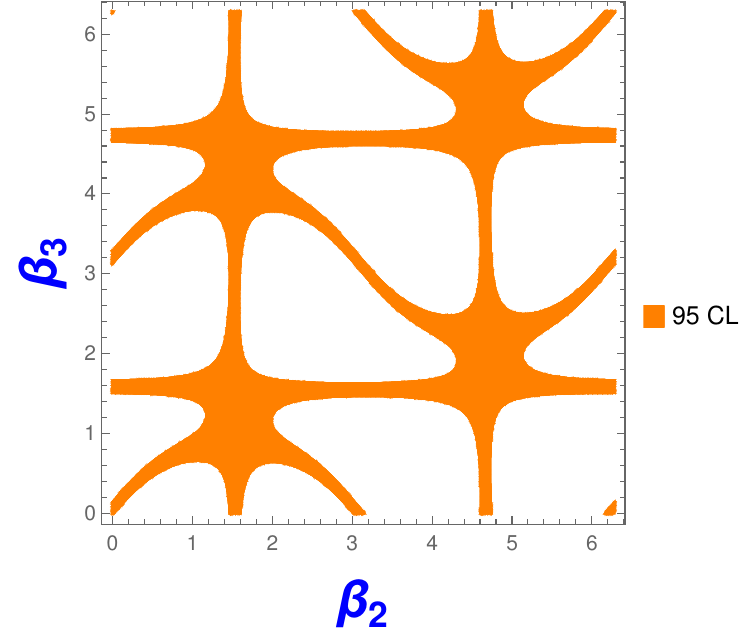}\\ 
\end{tabular}
\end{center}
\caption{
Colored regions correspond to the allowed parameter space of the angles $\beta_1$, $\beta_2$, and $\beta_3$, which parametrize the diagonalization of the mass matrix, at the 95\% C.L.. Their values depend on the Yukawa couplings and the VEVs of the Higgs doublets. The extent of the allowed region indicates the viability of the model. For a new physics scale of $1\,\text{TeV}$ (the three upper figures)  the model is essentially excluded, while at $6\,\text{TeV}$ (the other three figures) a non-negligible parameter space remains. The observables are listed in Table~\ref{tab:observables} and the corresponding pulls in Table~\ref{tab:pulls}.
}
\label{fig:95cl}	
\end{figure}
 \begin{table}
 \resizebox{0.98\textwidth}{!}{
 {\begin{tabular}{|c|c|c|c|}
 \hline  
 $\mathcal{O}$                    &Value~\cite{Qweak:2018tjf,ParticleDataGroup:2024cfk,Altmannshofer:2017fio}  & SM prediction $\mathcal{O}_{\text{SM}}$ ~\cite{ParticleDataGroup:2024cfk}       & $\Delta \mathcal{O}=\mathcal{O}-\mathcal{O}_{\text{SM}}$\\            
  \hline 
 $Q_W(p)$                         &$0.0719\pm 0.0045$       &$0.0708\pm 0.0003$ &  $4\left(\frac{M_{Z}}{g_{1}M_{Z'}}\right)^2 \Delta_A^{Eee}\left(2\Delta_{V}^{Uuu}+\Delta_{V}^{Ddd}\right) $\\            
  \hline 
 $Q_W(\text{\text{Cs}})$          &$-72.62\pm 0.43$       &$-73.25\pm 0.02$   & $Z\Delta Q_W(p)+N\Delta Q_W(n)$ \\
 \hline
 $Q_W(e)$                         &$-0.0403\pm 0.0053$    &$-0.0473\pm 0.0003$& $4\left(\frac{M_{Z}}{g_{1}M_{Z'}}\right)^2 \Delta_A^{ee}\Delta_{V}^{Eee}$\\
 \hline 
$1-\sum_{q=d,s,b}|V_{uq}|^2$      &$1-0.9999(6)$          &         0         & $\frac{3}{4\pi^2}\frac{M_W^2}{M_{Z'}^2}\left(\ln\frac{M_{Z'}^2}{M_W^2}\right)  \Delta_L^{E\mu\mu}\left(\Delta_L^{E\mu\mu}-\Delta_L^{Ddd}\right)    $      \\                               
\hline
  $\frac{\sigma^{\text{SM}+Z'}}{\sigma_{SM}}$& $0.83\pm 0.18$&       1         & $\frac{1+\left(1+4s_{W}^2+\Delta_V^{R\mu\mu}\Delta_L^{N\nu\nu}v^2/M_{Z'}^2\right)^2}{1+(1+4s_W^2)^2}-1$\\
\hline
$\frac{\Delta M_s^{\text{SM+NP}}}{M_s^{\text{SM}}}$& $\frac{17.757}{17.25}\pm 0.051$   & 1 & $\lvert 1+200\left(\frac{5\text{TeV}}{M_{Z'}}\right)^2\left[(\Delta_{L}^{Dsb})^2+(\Delta_{Rsb}^{Dsb})^2-\Delta_{L}^{Dsb}\Delta_{R}^{Dsb}\right]\rvert$-1\\
\hline
 \end{tabular}
 } 
}
 
\caption{ Experimental value and the new physics prediction for the shift in 
 the weak charge of the proton $Q_{W}(p)$~\cite{Qweak:2018tjf} (the neutron weak charge $Q_{W}(n)$ is similar to that of the proton by interchanging $u\leftrightarrow d$.), Cesium $Q_{W}(\text{Cs})$ and the electron $Q_{W}(e)$,
 owed to the interaction with the $Z'$. 
  The fourth observable is the constraint on the violation of the first-row CKM unitarity~\cite{Buras:2013dea,ParticleDataGroup:2024cfk}. We also impose the constraints on neutrino trident production~\cite{CCFR:1991lpl} reported by the CCFR collaboration. For the rotation from the weak basis
  to the mass eigenstates we adopt the convention~\cite{Barger:2003hg}: $\Delta^{ff}_{L,R}=g_{z'}\epsilon_{L,R}^{Z'}(f)$ 
  for up-type quarks, \textit{i.e.,}  $u$, $c$, $t$, right-handed down-type quarks, \textit{i.e.,} $d_R$, $s_R$, $b_R$~(to avoid FCNC) and charged leptons.   }
  \label{tab:observables}	
   \end{table}
  
A similar expression applies to neutrinos, but with the PMNS matrix, namely $\nu' = U^{N}\nu$ and $l' = U^{L}l$. In our framework, the lepton sector is universal, so lepton-flavor violation observables are not considered. Consequently, we adopt the standard choice $U^{N} = V^{PMNS}$ and $U^{L} = \mathbf{1}$, which is commonly used in this context~\cite{Langacker:2000ju}.
It is useful to define the vector and axial expressions $\Delta^{Fff}_{V,A}=\Delta^{Fff}_{R}\pm\Delta^{Fff}_{L}$~\cite{Buras:2012jb}.
In this expression, $F$ stands for a generic fermion, which may represent an up-type quark ($U$), a down-type quark ($D$), a neutrino-like lepton ($N$), or a charged lepton ($E$).

To express the contribution of new physics to low-energy observables, we define the $Z'$ contribution of the Hamiltonan in Eq.~(\ref{eq:331hnc}) for a given type of fermions $F$ as follows:
\begin{align}
 H^{F}_{Z'}=&  Z'_{\mu}
\sum_{f,f'}\bar{f}\gamma^{\mu}\left(\Delta_{L}^{Fff'}P_L+\Delta_{R}^{Fff'}P_R\right)f'\ .    
\end{align}

In Figure~\ref{fig:95cl} we illustrate the impact of the electroweak constraints listed in Table~\ref{tab:observables} on the parameter space of the three mixing angles that define the diagonalizing matrix connecting the interaction and mass basis. For a benchmark mass of $1\,\text{TeV}$, these constraints exclude most viable solutions. A consistent parameter space emerges only for $Z'$ masses greater than approximately $6\,\text{TeV}$. 
This requirement leads to bounds comparable to those obtained from collider searches, which set a lower limit on the $Z'$ mass of about $5\,\text{TeV}$ for this model~\cite{Benavides:2021pqx}, thus providing a complementary constraint.
To determine the allowed regions of the parameter space, we performed a $\chi^2$ analysis that incorporates the most relevant constraints on the $Z'$ sector.  
For models with nonvanishing axial couplings to the electron, i.e. $\epsilon_{L}^{Z'} - \epsilon_{R}^{Z'} \neq 0$,  
the most stringent bounds arise from parity-violation experiments. In particular, measurements of the weak charge of cesium~\cite{Patrignani:2016xqp,Wood:1997zq,Guena:2004sq} impose strong limits on the $Z'$ parameters.  
Competitive constraints are obtained from measurements of the weak charges of the proton and neutron~\cite{Androic:2018kni,Patrignani:2016xqp,Androic:2013rhu}, while determinations of the electron weak charge~\cite{Patrignani:2016xqp,Anthony:2005pm} provide complementary information since they involve only lepton couplings.  

An additional important constraint, depending solely on the left-handed chiral couplings of the muon and the down quark, arises from CKM unitarity~\cite{Marciano:1987ja,Buras:2013dea}.  
This bound is particularly relevant because it applies even in scenarios where the $Z'$ couplings to quarks vanish.  

We also include limits from neutrino trident production in the scattering of muon neutrinos off nuclei.  
This observable depends only on the couplings of the second lepton family and is therefore complementary to all the constraints discussed above.  
Its effect appears as corrections to neutrino--nucleon scattering, as summarized in the next-to-last row of Table~\ref{tab:observables}~\cite{Altmannshofer:2014cfa,Bian:2017rpg}.
As previously mentioned, this model is non-universal in the quark sector, and FCNC, together with collider searches, provide the most stringent constraints~\cite{DiLuzio:2019jyq}. In particular, we consider the restrictions from $B_s$--$\bar{B}_s$ mixing, expressed as  
\[
\frac{\Delta M_s^{\text{SM+NP}}}{\Delta M_s^{\text{SM}}}
   = \frac{17.7656 \pm 0.0057}{17.25 \pm 0.85}
   = 1.030 \pm 0.051 ,
\]  
where the numerator corresponds to the most recent LHCb combined measurement~\cite{LHCb:2021moh},  
$
\Delta M_s^{\text{exp}} = 17.7656 \pm 0.0057~\text{ps}^{-1}
$,  
and the denominator is the SM prediction from the 2023 PDG CKM fit, reported as either  
$
\Delta M_s^{\text{SM}} = 17.25 \pm 0.85~\text{ps}^{-1}
\quad \text{or} \quad 
\Delta M_s^{\text{SM}} = 16.54^{+0.50}_{-0.30}~\text{ps}^{-1}$.
From Fig.~\ref{fig:95cl} we observe that $Z'$ masses around $1$~TeV are essentially excluded, except for a set of parameters of nearly zero measure. Realizing such a scenario would require a fine-tuning of the model parameters in order to remain phenomenologically viable. As shown in Table~\ref{tab:pulls}, the best-fit point still provides an (almost) acceptable balance between $\chi^2_{\text{min}}$ and the number of degrees of freedom.  The fit is not optimal, reflecting a well-known difficulty of fitting a 3-3-1 model to low-energy data, because in this model most of the electroweak couplings are determined, and in our approximation, we can only vary the mixing angles of the diagonalization matrix.
For $M_{Z'} = 6$~TeV these constraints remain quite restrictive; nevertheless, in this case, there exists sufficient parameter space for a consistent model. Similar conclusions apply to other 3-3-1 realizations with non-universal couplings in the quark sector, so our analysis may be regarded as representative of this class of models.

\begin{table}[h!]
\centering
\begin{tabular}{|c|c|c|c|c|c|c|c|}
\hline
$\mathcal{O}$   &$Q_W(Cs)$&$Q_W(p)$&$Q_W(e)$&CKM  &$\nu$-Tri&$B_s$ mixing&$\chi^2/\text{n.o.f}$\\ \hline
pull$(m_{Z'}=6\text{TeV})$ & 1.400    & 0.205  & 1.386 &0.151& -0.945  & -0.539   & 5.13/(6-3)\\ \hline

pull$(m_{Z'}=1\text{TeV})$ & 0.276   & -0.383 & 1.768  &-0.080& -0.954  & -0.530   & 4.548/(6-3)\\ \hline
\end{tabular}
\caption{Pulls for each one of the observables $\mathcal{O}$ listed in Table~\ref{tab:observables}.
}
\label{tab:pulls}
\end{table}

\section{Conclusions}
In this work, we have constructed the full scalar potential of the 3-3-1 model, incorporating all interaction terms permitted by the gauge symmetries. The model was implemented in the \verb|SARAH| package, which allowed for the automatic generation of model files and provided an independent validation of our analytical results. Within this framework, we derived the mass matrices for neutral scalars, pseudoscalars, and charged scalars, and explicitly verified in specific cases that the correct number of Goldstone bosons is recovered, as required by the Nambu-Goldstone theorem. 

The inclusion of a scalar sextet plays a crucial role in simultaneously generating Dirac and Majorana mass terms. This feature makes it possible to obtain light neutrino masses through a Type-I seesaw mechanism. To connect with phenomenology, we further employed the Casas–Ibarra parametrization~\cite{Casas:2001sr} to extract the corresponding Yukawa couplings consistent with observed neutrino data.

We then focused on the most relevant regions of the parameter space, identifying the preferred mass ranges for the extended scalar spectrum. These results were presented in terms of probability density functions, which provide a statistical characterization of the model predictions. Remarkably, the resulting distributions show overlap with several experimental anomalies reported in diphoton final states, suggesting potential signals of the extended scalar sector.

We have also examined the electroweak constraints on the 3-3-1 model. In particular, we derived bounds on the $Z$–$Z'$ mixing angle as a function of the $Z'$ boson mass. In addition, we analyzed the constraints on the mixing angles of the quark mass diagonalization matrices, making use of the most recent data on weak charges of cesium, the proton, and the electron, as well as from neutrino trident production, CKM unitarity tests, and $B_s$–$\bar{B}_s^0$ mixing—the latter providing the strongest limits on flavor-changing neutral currents. We find that these indirect constraints are comparable in strength to those obtained from collider searches. Furthermore, since the 3-3-1 model with right-handed neutrinos is a well-motivated and widely studied benchmark within the class of 3-3-1 models, many of the conclusions from this phenomenological analysis are expected to apply more broadly to other 3-3-1 realizations without exotic electric charges.

Taken together, our results establish the 3-3-1 model with right-handed neutrinos as a viable and predictive framework, capable of simultaneously addressing the generation of neutrino masses, providing an extended scalar spectrum with distinctive collider signatures, and satisfying stringent low-energy constraints. The overlap of the predicted scalar mass distributions with reported anomalies in diphoton channels highlights its potential relevance for ongoing and future experimental programs. Further refinements, including the incorporation of additional collider and precision data, will sharpen the predictions and help assess whether the 3-3-1 framework can provide a consistent explanation for some of the observed deviations from the Standard Model.

\section{Acknowledgments}
RB and ER acknowledge financial support from Minciencias under grant CD82315 CT ICETEX 2021-1080. 
ER also acknowledges additional financial support from projects No.~3595 and No.~2679 of \emph{VIIS-UDENAR}.

\appendix


\section{SU(3) Invariants and the Scalar Potential \label{sec:su3invariants}}

\paragraph{Transformation properties of the sextet.}
In this section we will assume that, $\Phi_1$, $\Phi_2$ and $\Phi_3$ transform as antitriplets under 
$SU(3)_L$, i.e., $\Phi_{1,2,3}\sim 3$.
In this section we will asume that $\Phi_1$, $\Phi_2$ and $\Phi_3$ are triplets under 
$SU(3)_L$, i.e., $\Phi_{1,2.3}\sim 3$, 
while 
$\Phi_4$ transforms as a sextet $6$ with the same transformation properties as the tensor $\Phi_{1i}\otimes \Phi_{2j}$ (i.e., $3\otimes 3=\bar{3}_{a}+6$. 
Using the standard conventions for the contravariant and covariant indices in $SU(N)$, we define
\begin{align}
\Phi_{4}^{\prime ij}&= \tensor{U}{^i_k}\tensor{U}{^j_l}\Phi_{4}^{kl}, 
& \Phi_{4ij}^{* \,\prime}&= \tensor{U}{^*_i^k}\tensor{U}{^*_j^l}\Phi_{4kl}^{*}\ .
\end{align}
As usual, these transformations preserve the identity, which is equivalent to the invariance of the inner product $(\vec{a}^*,\vec{b})\to C$ for $\vec{a},\vec{b}\in \mathbb{C}^N$. Explicitly,
\begin{align}
\tensor{\delta}{^i_j}
= \tensor{U}{^i_k}\tensor{U}{^*_j^l}\tensor{\delta}{^k_l}
= \tensor{U}{^i_k}\tensor{U}{^*_j^k}
= \tensor{U}{^i_k}\tensor{U}{^{*T}^k_j}\ .    
\end{align}
In matrix notation, this is simply $\mathbf{1}=UU^{\dagger}$, and the sextet fields transform as
\begin{align}
\Phi_{4}^{\prime}= U\Phi_{4}U^T, 
&\hspace{0.5cm}
\Phi_{4}^{* \,\prime}= U^*\Phi_{4}^{*}U^{*T}\ .
\end{align}
Equivalently, 
\begin{align}
\Phi_{4}^{\dagger\prime}= U^T{}^{\!\dagger}\,\Phi_{4}^{\dagger}\,U^{\dagger}\ .
\end{align}
From this, it is immediate that $\text{Tr}[\Phi_4\Phi_4^{\dagger}]$ is invariant under $U(N)$.

\paragraph{Construction of the operator $\Phi_4\Phi_2^*\Phi_1\Phi_3$.}
The operator of interest arises in the tensor product
\footnote{For a complete list of tensor products, see~\cite{Slansky:1981yr}.}
\begin{align}
\underbrace{\Phi_4\Phi_2^*\Phi_1\Phi_3}_{6\otimes \bar{3}\otimes 3\otimes 3}\ .   
\end{align}
Expanding this product gives
\begin{align}
(6\otimes \bar{3})\otimes 3\otimes 3
&=(3+\bar{15})\otimes 3\otimes 3
=(\bar{3}+6_s+\cdots)\otimes 3
=1+\cdots\ .
\end{align}
From $6\otimes \bar{3}=3+\bar{15}$ we can form a $3$ by the contraction 
\begin{align}
\Phi_{4ik}\Phi_{2k}^* \sim 3_i\ .
\end{align}
Then, from $3\otimes 3=\bar{3}_a+6 $ we obtain a $\bar{3}$ by contracting
\begin{align*}
\bar{3} \sim \epsilon^{lij}3_i\Phi_{1j}
= \epsilon^{lij}(\Phi_{4ik}\Phi_{2k}^*)\Phi_{1j}\ .
\end{align*}
Finally, contracting this $\bar{3}$ with $\Phi_3$ (a $3$), we obtain a singlet:
\begin{align*}
\bar{3}\otimes 3\sim 
(\epsilon^{lij} \Phi_{4ik}\Phi_{2k}^*\Phi_{1j})\Phi_{3l}\ .
\end{align*}
Following the same procedure, we construct the remaining operators of the same structure:
\begin{align*}
&\lambda_{25} \epsilon^{ijl}  
 \Phi_{4ik}\Phi_{2k}^{*}\Phi_{1j}\Phi_{3l}
+\lambda_{26} \epsilon^{ijl}
\Phi_{4ik}\Phi_{2k}^{*}\Phi_{2j}\Phi_{3l}
+\lambda_{27} \epsilon^{ijl}
\Phi_{4ik}\Phi_{1k}^{*}\Phi_{1j}\Phi_{3l}
+\lambda_{28} \epsilon^{ijl}  
 \Phi_{4ik}\Phi_{1k}^{*}\Phi_{3j}\Phi_{2l}\ .
\end{align*}

\paragraph{Gauge invariance of the operator $\Phi_4\Phi_4\Phi_3\Phi_3$.}
To construct this operator we consider the product
\begin{align}
 (6\otimes 3)\otimes (6\otimes 3)
=(8+10)\otimes (8+10)
=8\otimes 8+\cdots= 1+\cdots\ .    
\end{align}
To demonstrate singlet gauge invariance, it is convenient to examine how the octet $8$ transforms under $SU(3)$:
\begin{align}
8'=\tensor{A}{^'^l_i}
&=\epsilon_{ijk}\Phi_{3}^{j\prime}\Phi_{4}^{kl\prime}
=\epsilon_{ijk}\tensor{U}{^j_{k'}}\Phi_{3}^{k'}
\tensor{U}{^k_m}\tensor{U}{^l_n}\Phi_{4}^{mn}\notag\\
=&
\epsilon_{ojk}\tensor{U}{^j_{k'}}
\underbrace{[\tensor{U}{^o_p}\tensor{U}{^{\dagger}^p_i}]}_{\delta^{oi}}
\Phi_{3}^{k'}\tensor{U}{^k_m}\tensor{U}{^l_n}\Phi_{4}^{mn}\notag\\
=& 
(\epsilon_{ojk}
\tensor{U}{^o_p}
\tensor{U}{^j_{k'}}
\tensor{U}{^k_m})
\tensor{U}{^{\dagger}^p_i}
\Phi_{3}^{k'}\tensor{U}{^l_n}\Phi_{4}^{mn}\notag\\
=& 
(\epsilon_{pk'm}\det(U))
\tensor{U}{^{\dagger}^p_i}
\tensor{U}{^l_n}\Phi_{3}^{k'}\Phi_{4}^{mn}
=
\epsilon_{pk'm}
\tensor{U}{^{\dagger}^p_i}
\tensor{U}{^l_n}\Phi_{3}^{k'}\Phi_{4}^{mn}\ .
\end{align}
Where we take into account that $\epsilon_{ijk}\tensor{U}{^i_m}\tensor{U}{^j_n}\tensor{U}{^k_{l}}= \epsilon_{mnl}\text{det}U $ 
and  $\det U=1$ for $SU(3)$, one finds
\begin{align}
\tensor{A}{^'^l_i}
=(\epsilon_{pk'm})\tensor{U}{^{\dagger}^p_i}
\tensor{U}{^l_n}\Phi_{3}^{k'}\Phi_{4}^{mn}\ .
\end{align}
Thus, $\tensor{A}{^i_l}$ transforms as an octet, and the invariant singlet is obtained by contracting all indices of its square:
\begin{align}\label{eq:AA1}
\tensor{A}{^'^l_i}\tensor{A}{^'^i_l}
&=\epsilon_{pk'm}\tensor{U}{^{\dagger}^p_i}\tensor{U}{^l_n}\Phi_{3}^{k'}\Phi_{4}^{mn}
\;\epsilon_{p'k''m'}\tensor{U}{^{\dagger}^{p'}_l}\tensor{U}{^i_{n'}}\Phi_{3}^{k''}\Phi_{4}^{m'n'} \ .
\end{align}
Using
\begin{align}
  \tensor{U}{^{\dagger}^p_i}\tensor{U}{^i_{n'}}=\tensor{\delta}{^p_{n'}}\ , 
&\hspace{1cm}
\tensor{U}{^l_n}\tensor{U}{^{\dagger}^{p'}_l}=\tensor{\delta}{_n^{p'}}\ ,
\end{align}
the expression~\eqref{eq:AA1} reduces to
\begin{align*}
\tensor{A}{^n_p}\tensor{A}{^p_n}
&=\epsilon_{pk'm}\Phi_{3}^{k'}\Phi_{4}^{mn}
\;\epsilon_{nk''m'}\Phi_{3}^{k''}\Phi_{4}^{m'p} \ ,
\end{align*}
Thus, demonstrating that this expression is gauge invariant. 

\section{Diagonalizing a General $3\times 3$ Symmetric Matrix}\label{sec:appendixDiagonal}
In this appendix we provide an analytical expression for a matrix that diagonalize a general symmetric matrix $m_R$ with unknown elements, in our case a Majorana mass matrix. Starting from the diagonalization condition:

\begin{equation}\label{eq:diag}
    R^\dagger \,m_R \,R =\text{diag}~\{m_1, m_2, m_3\}\,,
\end{equation}
where $R$ is a unitary matrix and $m_i$ are the $m_R$ eigenvalues, in our case proportional to the heavy neutrino masses. It is then straighforward to write {\it diagonalization invariants}, i.e, the determinant (Det) and the trace of any power of $m_R$, which applied to Eq.~(\Ref{eq:diag}) can be written in the form:

\begin{equation}\label{eq:invariants}
    \begin{split}
        \text{Det}\{m_R\}& =\text{Det}\{\text{diag}\{m_1, m_2, m_3\}\}\,,\\
         \text{Trace}\{m_R\}& =\text{Trace}\{\text{diag}\{m_1, m_2, m_3\}\}\,,\\
          \text{Trace}\{m_R^2\}& =\text{Trace}\{\text{diag}\{m_1^2, m_2^2, m_3^2\}\}\,.
    \end{split}
\end{equation}
The invariants in Eq.~(\ref{eq:invariants}) relate the $m_R$ eigenvalues $m_i$, assumed to be known, with the $m_R$ elements $m_{ij}$ that satisfy Eq.~(\ref{eq:diag}), and for any unitary matrix $R$. One can therefore find some $m_{ij}$ elements in terms of the eigenvalues by solving the non-linear system that results from Eq.~(\ref{eq:invariants}). Notice however that a real $3\times 3$ symmetric matrix have six real parameters and we have only three equations, and therefore there is no a unique solution. We have additionally assumed non-negative eigenvalues $m_i>0$ (as a physical condition in our case) and selected a solution with the least number of free parameters, as explicitly given below. After replacing the obtained $m_{ij}$ elements in $m_R$ one can use Eq.~(\Ref{eq:diag}) to find the $R_{ij}$ elements . \\

Defining the following dimensionless factors:

\begin{equation}\label{eq:factors}
\begin{split}
    f_1 &\equiv \frac{2 m_{23}^2  + ( m_{33}-m_2) (m_{33}-m_3)}{(m_1 - m_2) (m_1 - m_3)}\\
    f_2 &\equiv  -\frac{2 m_{23}^2  + (m_{33}-m_1) (m_{33}-m_3 )}{(m_1 - m_2) (m_2 - m_3)}\\
    f_3 &\equiv \frac{2 m_{23}^2  + (m_{33}-m_1) (m_{33}-m_2)}{(m_1 - m_3) (m_2 - m_3)}\,,\\
\end{split}
\end{equation}
and also, in order to write simpler expressions for each element the diagonalizing matrix $R$, additional dimensionful quantities are defined. In particular:

\begin{equation}\label{eq:rootfactor}
\begin{split}
    F\equiv -m_{23}^4 & \left[2 m_{23}^2  + (m_{33}-m_1 ) (m_{33}-m_2) \right] \times \\
      &\left[2 m_{23}^2  + (m_{33}-m_1) ( m_{33}-m_3)\right] \times \\ 
      &\left[2 m_{23}^2  + (m_{33}-m_2) (m_{33}-m_3)\right]
\end{split}
\end{equation}

\begin{equation}
\begin{split}
    n_{11} &\equiv -2 m_{23}^6  - m_{23}^4  (m_{33}-m_1) (m_1 - m_2 - m_3 + m_{33})\\
    d_{11} &\equiv-m_{23}^2  (m_{33}-m_1 ) \left[2 m_{23}^2  + ( m_{33}-m_2) (m_{33}-m_3)\right]
\end{split}
\end{equation}

\begin{equation}
R_{11}=\sqrt{f_1}  \frac{(-m_{33} + m_1)\sqrt{F} + n_{11}}{m_{23}(\sqrt{F} + d_{11})}
\end{equation}

\begin{equation}
\begin{split}
 n_{12} \equiv 2 m_{23}^5  &+ m_{23} (m_{33}-m_1 )^2  (m_{33}-m_2 ) (m_{33}-m_3) \\
 &+ m_{23}^3  ( m_{33}-m_1) (-m_1 - m_2 - m_3 + 3 m_{33})
\end{split}
\end{equation}

\begin{equation}
\begin{split}
    R_{12}&=\sqrt{f_1} (n_{12}/(\sqrt{F}+d_{11}))\\
    R_{13}&=\sqrt{f_1}
\end{split}
\end{equation}

\begin{equation}
\begin{split}
 n_{21} & \equiv -2 m_{23}^6  - m_{23}^4  (m_{33}-m_2) (-m_1 + m_2 - m_3 + m_{33})\\
d_{21} & \equiv -m_{23}^2  (m_{33}-m_2) \left[2 m_{23}^2  + (m_{33}-m_1) (m_{33}-m_3)\right]
\end{split}
\end{equation}

\begin{equation}
    R_{21}=\sqrt{f_2}  \frac{(-m_{33} + m_2)\sqrt{F} + n_{21}}{m_{23} (\sqrt{F}+d_{21})}
\end{equation}

\begin{equation}
\begin{split}
    n_{22} \equiv 2 m_{23}^5  &+ m_{23} (m_{33}-m_1) (m_{33}-m_2)^2  (m_{33}-m_3) \\ 
   &+ m_{23}^3  (m_{33}-m_2) (-m_1 - m_2 - m_3 + 3 m_{33})
\end{split}
\end{equation}

\begin{equation}
\begin{split}
    R_{22}&=\sqrt{f_2}  \frac{n_{22}}{\sqrt{F}+d_{21}}\\
    R_{23}&=\sqrt{f_2}
\end{split}
\end{equation}

\begin{equation}
\begin{split}
    n_{31} &\equiv -2 m_{23}^6  - m_{23}^4  (m_{33}-m_3) (-m_1 - m_2 + m_3 + m_{33}) \\
d_{31} & \equiv m_{23}^2  (m_3 - m_{33}) \left[2 m_{23}^2  + (m_{33}-m_1) (m_{33}-m_2)\right]
\end{split}
\end{equation}

\begin{equation}
    R_{31}=\sqrt{f_3}  \frac{(-m_{33} + m_3) \sqrt{F}  + n_{31}}{m_{23}(\sqrt{F} + d_{31})}
\end{equation}

\begin{equation}
\begin{split}
    n_{32} \equiv 2 m_{23}^5  &+ m_{23} ( m_{33}-m_1) (m_{33}-m_2) (m_{33}-m_3)^2 \\
   & +m_{23}^3  (m_{33}-m_3) (-m_1 - m_2 - m_3 + 3 m_{33})
\end{split}
\end{equation}

\begin{equation}
\begin{split}
        R_{32}&=\sqrt{f_3}  \frac{n_{32}}{\sqrt{F}+d_{31}}\\
    R_{33}&=\sqrt{f_3} \,.
\end{split}
\end{equation}

The chosen solution of the system in Eq.~(\ref{eq:invariants}), from which the $R$ matrix was obtained, depends on two free parameters, i.e, $m_{23}$ and $m_{33}$, two elements of the symmetric matrix $m_R$. Each of the $R$ elements are given in terms of this two elements and the eigenvalues $m_i$. From Eq.~(\ref{eq:factors}) one can see that the $m_R$ eigenvalues cannot be degenerated, $m_1\ne m_2 \ne m_3$. In our specific case, we have assumed the heavy neutrino masses in a hierarquical order, i.e, $m_1<m_2<m_3$

To obtain finite $R$ matrix, additional mathematical conditions must be fulfilled, basically:

\begin{equation}
    \sqrt{F}+d_{11} \ne 0 \quad \& \quad \sqrt{F}+d_{21} \ne 0 \quad  \& \quad \sqrt{F}+d_{31} \ne 0 \,,
\end{equation}
with $m_{23}\ne 0$ and where $f$ is defined in Eq.~(\ref{eq:rootfactor}). 

Finally, requiring all the $R$ matrix elements be real, imply the following mathematical conditions:

\begin{equation}\label{eq:mathConditions}
    f_1 \ge 0 \quad \& \quad f_2 \ge 0 \quad  \& \quad f_3 \ge 0 \,,
\end{equation}
where $f_i$'s are defined in Eq.~(\ref{eq:factors}). This last requirements will reduce the range of the $m_{23}$ and $m_{33}$ elements, which will be bounded only by relations of the $m_R$ eigenvalues. Notice that the factor $F$ in Eq.~(\ref{eq:rootfactor}) and the $d_{i1}$ definitions are related with the $f_i$ factors in Eq.~(\ref{eq:factors}), and therefore only fullfilling the conditions in Eq.~(\ref{eq:mathConditions}) was enough for our purpose, i.e, a finite $R$ real matrix is obtained.


\bibliographystyle{unsrt}
\bibliography{referencias}

\end{document}